\documentclass[aps,pre,reprint,groupedaddress,showkeys,showpacs]{revtex4-1}

\usepackage[dvipdfm]{hyperref}
\usepackage[dvipdfm]{graphicx}
\usepackage{longtable}
\usepackage{url}
\usepackage{bm}
\usepackage{amsmath}
\usepackage{color}
\begin{document}


\title{Simulated Tempering and Magnetizing: An Application of Two-Dimensional Simulated Tempering to Two-Dimensional Ising Model and Its Crossover}


\author{Tetsuro NAGAI$^1$ and Yuko OKAMOTO$^{1,2,3}$}
\affiliation{$^1$Department of Physics, Graduate School of Science, Nagoya University, Nagoya, Aichi 464-8602, Japan \\
	  $^2$Structural Biology Research Center, Graduate School of Science, Nagoya University, Nagoya, Aichi 464-8602, Japan\\
	  $^3$Center for Computational Science, Graduate School of Engineering, Nagoya University, Nagoya, Aichi 464-8603, Japan}



\begin{abstract}
We performed two-dimensional simulated tempering (ST) simulations of the two-dimensional Ising model with 
different lattice sizes in order to investigate the two-dimensional ST's applicability 
to dealing with phase transitions and to study the crossover of critical scaling behavior.
The external field, as well as the temperature, was treated as a dynamical variable updated during the simulations.  
Thus, this simulation can be referred to as ``Simulated Tempering and Magnetizing (STM)." 
We also performed the ``Simulated Magnetizing" (SM) simulations,  in which the external field was 
considered as a dynamical variable  and temperature was not. 
As has been discussed by previous studies, the ST method is not always compatible with first-order phase transitions. 
This is also true in the magnetizing process. Flipping of the entire magnetization did not occur
in the SM simulations under $T_\mathrm{c}$ in large lattice-size simulations. 
However, the phase changed through the high temperature region in the STM simulations.
Thus, the dimensional extension let us eliminate the difficulty of the first-order phase transitions and study wide
area of the phase space.  
We then discuss how frequently parameter-updating attempts should be made for optimal convergence.
The results favor frequent attempts.  
We finally study the crossover behavior of the phase transitions with respect to the temperature and external field. 
The crossover behavior was clearly observed in the simulations in agreement with the theoretical implications.

\end{abstract}


\maketitle


\section{Introduction}

In the computational statistical physics field,  Monte Carlo (MC) and molecular dynamics (MD)  simulations have been commonly used. 
However, the quasi-ergodicity problem, where simulations tend to get trapped 
in states of energy local-minimum, has often posed a great difficulty. 
In order to overcome this difficulty, generalized-ensemble algorithms 
have been developed and applied to many systems 
including spin systems and biomolecular systems (for reviews, see, e.g., Refs.\ \cite{Hansmann1999,Mitsutake2001,Sugita2001}).

Commonly used examples of generalized-ensemble algorithms are  multicanonical (MUCA) 
\cite{berg1991multicanonical,berg1992multicanonical}, 
simulated tempering (ST) method \cite{Lyubartsev1992,marinari1992simulated}, 
and replica-exchange method (REM) \cite{hukushima1996exchange,Geyer1991} (it is also referred to as parallel tempering). 
Closely related to MUCA are the Wang-Landau method \cite{Wang2001a,Wang2001b} and metadynamics \cite{Laio2002}.
Also closely related to REM is the method in Ref.\ \cite{Swendsen1986}.

In the ST method, temperature is regarded as a dynamical variable, which is updated by the Metropolis criteria during the simulation, and consequently
a random walk is realized in the temperature space. This random walk, in turn, causes a random walk of the energy, which enables the 
system in question to overcome free-energy barriers.  However, it is well-known that the ST method is
not very compatible with first-order phase transitions (for a review, see, e.g., Ref.\ \cite{iba2001extended}).  
When there is a first-order phase transition,
the random walk of temperature across the phase-transition point hardly occurs. 
We remark that there is a recent attempt to deal with this difficulty by
an extension of ST \cite{kim2010}.

Recently, the multi-dimensional generalizations of 
the generalized-ensemble algorithms, including MUCA, ST, and REM,  were discussed and general formalisms were given 
\cite{Mitsutake2009multidimensional1,Mitsutake2009multidimensional2,Mitsutake2009MSTMREM}.  
In these methods, the energy of the system is generalized by adding other energy term(s) with some coupling constants.
In the multi-dimensional ST method, not only the temperature but also the coupling constants are considered as dynamical variables. 

In this work, we study a special case of the  above general multi-dimensional ST methods. 
Namely, the additional term is $-hM$  where $h$ and $M$ are the external field and the magnetization, respectively.
The external field $h$ corresponds to the couping constant
which is updated during MC simulations. Therefore, not only temperature but also external field becomes a dynamical variable 
and is expected to realize a random walk during the simulations. 
Thus, this simulation can be referred to as the ``Simulated Tempering and Magnetizing" (STM).
In order to test the effectiveness of the present method, we applied it to the two-dimensional Ising model. 

The Ising model has two  kinds of phase transitions. One occurs along the change of temperature when the external field is zero. 
The other occurs along the change of external field when the temperature is under the critical temperature ($T_\mathrm{c}$). 
The former is classified into a second-order phase transition.
The latter is categorized as a first-order phase transition unless the temperature is exactly equal to  $T_\mathrm{c}$. 
When $T=T_\mathrm{c}$, the transitions are classified into a second-order phase transition.
This system allows us to confirm applicability of the two-dimensional ST 
to the first-order phase transitions along the external field changes.

We also investigated the crossover phenomena in the phase transitions, in which critical exponents are changed. 
We study the behavior of magnetization per spin $m$, 
which follows $m \sim |T-T_\mathrm{c}|^{\beta}$ and $m \sim |h|^{1/\delta}$ near the critical point, 
where $\beta$ and $\delta$ are critical exponents \cite{gaunt1970equation}. 
Our simulation method, with a combination of histogram reweighting techniques, enables us to calculate physical values 
such as the energy and magnetization at various values of $T$ and $h$ from a single production run.

This article is organized as follows. In \S 2 we present the STM method. 
In \S 3 we present the results. In \S 4 we conclude this article.

\section{Materials and Methods}

\subsection{System}
We study the two-dimensional Ising model in external field.
The total energy is given by
\begin{align}
H&=E-hM\,,\\
E&=-\sum_{\left< i,j\right>} \sigma_i \sigma_j \label{eq:E}\,,\\
M&=\sum_{i=1}^N \sigma_i \,,\label{eq:M}
\end{align}
where $i$, $N$, $\sigma_i$, and $h$ are the index of spin, total number of spins,
spin at the $i$-th site, and external field, respectively. The spin $\sigma_i$ takes on the values $\pm1$.
The sum in Eq.\ (\ref{eq:E}) goes over the nearest-neighbor pairs. 
The spins are arranged on the square $L\times L$ lattice. 
We imposed the periodic boundary conditions. 
Data were obtained for lattice sizes from $2\times2$ to $160\times 160$.

\subsection{Simulation methods}
Whereas the conventional ST method considers temperature as a dynamical variable,
the STM method considers not only temperature but also external field as a dynamical variable.
Here, before explaining the STM method, we shortly review the conventional ST method \cite{marinari1992simulated,Lyubartsev1992}.

In the conventional ST method, temperature is a dynamical variable  
which takes on one of  $N_T$ values  (here, temperature is discretized into $N_T$ values).
In other words, denoting $X$ and $x$ as a sampling space and its microscopic state, respectively, 
the Boltzmann factor 
\begin{align}
e^{-E(x)/T +a(T)}
\end{align} 
is regarded as a joint probability for the state $(x,T)$ ($\in X \otimes \{T_1, T_2, \dots , T_{N_T}\}$,    
the product space of $X$ and $\{T_1, T_2, \dots , T_{N_T}\}$).  
Here, $a(T)$ (or $a(T_i)$) is a parameter for obtaining uniform distributions of temperature values.  
Here and hereafter, we set Boltzmann's constant to unity.
Now that the temperature is a dynamical variable, 
the simulated system is allowed to realize a random walk in the temperature space.
This random walk, in turn, causes a random walk of energy. 
Consequently the simulated system has more chance to overcome energy barriers. 
      
Even though temperature changes during ST simulations, 
any thermodynamic quantity at temperature $T_i$, $\left< A\right> _{T_i}$, can be reconstructed with 
the conditional expectation of a physical quantity $A$ given at $T_i$, or $\left<A|T_i\right>$. Note that 
\begin{align}
    \left< A|T_i\right>_{ST}  
    &=\frac{\displaystyle\sum _{j=1} ^{N_T} \int dx A(x) \delta_{ij} e^{-\frac{E(x)}{ T_j} +a(T_j)}}
        {\displaystyle\sum _{j=1} ^{N_T} \int dx  \delta_{ij} e^{-\frac{E(x)}{T_j} +a(T_j)}}\\
    &= \frac{ \displaystyle\int dx A(x) e^{-\frac{E(x)}{ T_i} +a(T_i)}}{\displaystyle\int dx  e^{-\frac{E(x)}{T_i} +a(T_i)}}\\
    &= \left< A \right>_{T_i}  \,, 
\end{align}
where $\delta_{ij}$ is the Kronecker delta. 
Namely, we have  
\begin{align}
    \left< A\right>_{T_i} = \frac{1}{N_{T_i}}\sum_{j=1}^{N_{T_i}} A^{j}_{T_i} \,,  
\end{align}
where $N_{T_i}$ and $A^{j}_{T_i}$ stand for the total number of samples  and  $j$-th sample at $T_i$.

To find the candidate for $a(T_i)$, let us look into the probability of visiting $T_i$. 
By summing over the delta function, the probability of occupying $T_i$ is given by 
\begin{align}
P(T_i) &= \frac{\displaystyle\sum _{j=1} ^{N_T} \int dx \delta_{ij} e^{-\frac{E(x)}{ T_j} +a(T_j)}}
            {\displaystyle\sum _{j=1} ^{N_T} \int dx e^{-\frac{E(x)}{ T_j} +a(T_j)}}\\
       &= \frac{e^{-f(T_i) +a(T_i)} }{\displaystyle\sum _{j=1} ^{N_T} e^{-f(T_j) +a(T_j)}} \\
       &\propto e^{-f(T_i) +a(T_i)} \,,
\end{align}
where $f$ is the dimensionless (Helmholtz) free energy and 
\begin{align}
e^{-f(T)} \equiv \int dx e^{-E(x)/T}.
\end{align}
Substituting $f(T_i)$  into $a(T_i)$ gives constant probability regardless of $T_i$.
Thus, the dimensionless free energy $f(T_i)$ is a good choice for $a(T_i)$ in order to obtain uniform temperature distribution 
and  to realize a random walk in the temperature space.
Although the free energy is not known {\em a priori}, unless the system is exactly solvable,
the free energy calculation methods (the details will be provided below) enable us to get its good estimate from preliminary simulation runs.

In the two-dimensional ST algorithm, on the other hand, 
we consider that another parameter is also a dynamical variable 
\cite{Mitsutake2009MSTMREM,Mitsutake2009multidimensional2,Mitsutake2009multidimensional1}. 
Especially in the STM method, the external field $h$ is a second dynamical variable.  
In other words, we consider 
\begin{align}
e^{-(E-hM)/T +a(T,h)} \label{eq:f1}
\end{align}
as a joint probability for $(x,T,h)$ ($\in X \otimes \{ T_1, T_2, \dots, T_{N_T}\} \otimes \{h_1, h_2, \dots h_{N_h}\}$), 
where $a(T,h)$ is a parameter. 

To find the candidate for $a(T_i,h_j)$, we again look into 
the probability of staying at each set of parameter values.  It is given by
\begin{widetext}
\begin{align}
P(T_i,h_j) &= 
        \frac{\displaystyle\sum _{k=1} ^{N_T}\displaystyle\sum _{l=1} ^{N_h} 
                       \int dx \delta_{ik} \delta_{jl} e^{-\frac{E(x)-h_l M(x)}{T_k} +a(T_k, h_l)} }
             {\displaystyle\sum _{k=1} ^{N_T}\displaystyle\sum _{l=1} ^{N_h} \int dx  e^{-\frac{E(x)-h_l M(x)}{T_k} +a(T_k, h_l)}}\\
           &=
              \frac{e^{-f(T_i,h_j)+a(T_i, h_j)} }
                   {\displaystyle\sum _{k=1} ^{N_T}\displaystyle\sum _{l=1} ^{N_h}  e^{-f(T_k,h_l) +a(T_k, h_l)}} \\
           &\propto e^{-f(T_i,h_j)+a(T_i, h_j)} \,, \label{eq:P}
\end{align}
\end{widetext}
where 
\begin{align}
e^{-f(T_i,h_j)}=\int dx e^{-(E-h_jM)/T_i}. \label{eq:f2}
\end{align}
The dimensionless free energy $f(T_i,h_j)$ is again a good choice 
for $a(T_i,h_j)$ in order to acquire uniform distribution of $T$ and $h$.
These values can be estimated from preliminary simulation runs and reweighting techniques.   

As in conventional ST method, any thermal average $\left< A\right> _{T_i,h_j}$ 
at given $T_i$ ($\in \{T_1,T_2,\dots,T_{N_T}\}$) and $h_j$ ($\in \{h_1,h_2,\dots,h_{N_h}\}$) 
can be obtained by calculating the conditional expectation: 
$ \left< A\right>_{T_i, h_j} = \left<A|T_i, h_j\right>_{ST} $. Namely, we have 
\begin{align}
\left< A\right>_{T_i,h_j} = \frac{1}{N_{T_i,h_j}}\sum _{k=1}^{N_{T_i,h_j}} A^{k}_{T_i,h_j} \,,
\end{align}
where $N_{T_i,h_j}$ is the total number of samples at $T_i$ and $h_j$, and $A^{k}_{T_i,h_j}$ stands for the $k$-th sample at $T_i$ and $h_j$.

The method of updating $T$ or $h$ is similar to that of updating spins because $T$ and $h$ are considered as dynamical variables.
The Metropolis criterion for updating $T$ or $h$ is given by the following transition probability:
\begin{widetext}
\begin{align}
w(T_i, h_j &\rightarrow  T_{i'}, h_{j'}) = \min  \left( 1, \frac{P(T_{i'}, h_{j'})}{P(T_i, h_j)}\right)\\
                                   &= \min  \left(1, \exp\left(-\left(\frac{1}{T_{i'}}-\frac{1}{T_i}\right) E+
\left(\frac{h_{j'}}{T_{i'}}-\frac{h_j}{T_i} \right) M + a(T_{i'},h_{j'}) -a(T_i,h_j)\right)\right) \,.
\label{eq:trans}
\end{align}
\end{widetext}

Once an initial state is given, the STM simulations can be performed by repeating the following two steps.  
1.\ We perform a conventional canonical simulation at $T_i$ and $h_j$ for certain MC sweeps. 
2.\ We update the temperature or external field by Eq.\ (\ref{eq:trans}) with $a(T,h)=f(T,h)$.

In our implementation every certain MC sweeps either $T$ or $h$  was updated 
 (the choice between $T$ and $h$ was made at random)
by Eq.\ (\ref{eq:trans}) to a neighboring value (the choice of two neighbors was also made at random). 
Here, one MC sweep stands for $L\times L$ single spin updates.
The number of MC sweeps performed between parameter updates is here  referred to as the parameter-updating period. 

Whereas updating the parameter to a neighboring value with the Metropolis algorithm should be 
considered the easiest to implement, 
we remark that, as spins can be updated by a number of methods such as the heat bath method, 
other schemes of updating the parameters can be employed \cite{chodera2011replica}. 
There also exists a temperature updating scheme for ST by Langevin algorithm \cite{zhang2010}.

Table \ref{table} summarizes the  conditions of the present simulations. For $L=80$, instead of   
a single 4000000000 MC sweep production run, four 1000000000 MC sweep runs were performed.
This was just to make one trajectory shorter and easier to deal with numerically.
Similarly, two production runs (instead of a single run) were made for $L=30$ and $160$.

\begin{table*}
\caption{Conditions of the two-dimensional ST simulations. }
\label{table} 
\begin{ruledtabular}
\begin{tabular}{lccccc}
Lattice size   $L$   &  2, 4, 8, 10, 20&       30&       80&   160       \\ \hline
Number of production runs &              1&        2&        4&     2       \\       
Total MC sweeps per run&       42000000& 42000000&  1000000000& 321300000\\
Parameter-updating period&         50  &       20&          10&         5\\ 
$T_1$--$T_{N_T}$       &       1.0--5.0& 1.0--5.0&   1.0--5.0 &  1.0--3.6\\
$h_1$--$h_{N_h}$       &      -1.5--1.5&-1.5--1.5&   -1.5--1.5& -0.5--0.5\\
$N_T$                  &         20    &   20    &          70&        63\\
$N_h$                  &         21    &   21    &          51&        51\\
$N_{{\rm data}}$\footnote{The data were stored every $N_{{\rm data}}$ MC sweeps.}        &         10    &   10    &         100&        50\\
\end{tabular}
\end{ruledtabular}
\end{table*}

As for spin-updates, we employed the single spin update algorithm; we updated spins one by one with the Metropolis criteria.
As for quasi-random-number generator, we used the Mersenne Twister \cite{matsumoto1998mersenne}.

\subsection{Free energy calculations}
The simulated tempering parameters, or free energy in Eqs.\ (\ref{eq:f1}) and (\ref{eq:f2}) can be simply obtained 
by the reweighting techniques applied to the results of
preliminary simulation runs 
\cite{mitsutake2000replica,Mitsutake2009MSTMREM,Mitsutake2009multidimensional2,Mitsutake2009multidimensional1}. 
We employed two reweighting methods for this free energy calculation. 
One method is the multiple-histogram reweighting method, or Weighted Histogram Analysis Method (WHAM) \cite{Ferrenberg1989,kumar1992weighted} 
and the other is Multistate Bennett Acceptance Ratio estimator (MBAR) 
\cite{shirts2008statistically}, which is based on WHAM. 

The equations of WHAM algorithm applied to the system is as follows. For details, 
the reader is referred to Refs.\ \cite{kumar1992weighted, Mitsutake2009multidimensional2}.
The density of states (DOS) $n(E,M)$ and free energy values $f(T_i,h_j)$ can be obtained by 
\begin{widetext}
\begin{align}
n(E,M) &= \frac{\displaystyle\sum _{T_i, h_j} n_{T_i,h_j}(E,M)}
                {\displaystyle\sum_{T_i,h_j} N_{T_i,h_j} \exp 
(f(T_i,h_j)-(E-h_jM)/T_i)} \,,  \label{eq:wham1}\\
f(T_i,h_j) &= -\log \sum_{E,M} n(E,M) \exp (-(E-h_jM)/T_i)   \,, \label{eq:wham2}
\end{align}
\end{widetext}
where $n_{T_i,h_j}(E,M)$ is the histogram of $E$ and $M$ at $T_i$ and 
$h_j$, and $N_{T_i, h_j}$ is the total number of 
samples obtained at $T_i$ and $h_j$.
By solving these two equations self-consistently by iterations, we can 
obtain $n(E,M)$ and $f(T_i, h_j)$.
The obtained $n(E,M)$ allows one to calculate any thermal average at arbitrary temperature and external field values.
Note that $f(T_i,h_j)$ are determined up to a constant, which set the zero point of free energy. 
Accordingly, $n(E,M)$ are determined up to a normalization constant.

The MBAR is based on the  following  equations. Namely, by combing Eqs. (\ref{eq:wham1}) and (\ref{eq:wham2}), the free energy can be written as  
\begin{widetext}
\begin{align}
f(T_i,h_j) = - \log \displaystyle\sum_{i=1}^{N_T}\sum_{j=1}^{N_h}\sum_{n=1}^N 
  \frac{ \exp (-(E_n-h_j M_n)/T_i)}
       {\displaystyle\sum_{k=1}^{N_T}\displaystyle\sum_{l=1}^{N_h}N_{T_k,h_l}
\exp (f(T_k,h_l)-(E_n-h_lM_n)/T_k)} \,,
\end{align}
\end{widetext}
where $N$, $N_{T_k, h_l}$, $E_n$, and $M_n$ is the total number of data, 
the number of samples associated with $T_k$ and $h_l$, energy of the $n$-th data, and magnetization 
of the $n$-th data, respectively.
This equation should be solved self-consistently for $f(T_i,h_j)$. 
Note that, as in WHAM, $f(T_i,h_j)$ are determined up to a constant.

We repeat the preliminary STM simulations and free energy calculations 
until we finally obtain sufficiently accurate free energy values 
which let the system perform a random walk in the temperature and external field space during the STM simulation. 
We then perform a single, final production run. 

Note that these two reweighting methods enable us to obtain not only dimensionless free energy values 
but also physical values at any temperature and at any external field.
It is given by 
\begin{widetext}
\begin{align}
\left< A \right> _{T,h} &= \sum_{n=1}^{N} W_{na} A(x_n) \,, \\
W_{na} &= \frac{1}{\left< c_a \right>} \frac{\exp (-(E_n-hM_n)/T)}
            {\displaystyle\sum_{k=1}^{N_T}\displaystyle\sum_{l=1}^{N_h} 
N_{T_k,h_l} \exp (f(T_k,h_l)-(E_n-h_l M_n)/T_k)} \,, \\
        \left< c_a \right> &= \displaystyle\sum_{n=1}^{N} 
            \frac{\exp (-(E_n-hM_n)/T)}
                 {\displaystyle\sum_{k=1}^{N_T}\displaystyle\sum_{l=1}^{N_h} N_{T_k,h_l} \exp (f(T_k,h_l)-(E_n-h_l M_n)/T_k)} \,.
\end{align}
\end{widetext}
For  details, the reader is referred to Refs.\ \cite{shirts2008statistically,Mitsutake2003}.

We also used another method of calculating free energy.
By substituting $a(T,h)$ in Eq.\ (\ref{eq:P}) by the estimates for free energy $\tilde{f}(T,h)$, we obtain
\begin{align}
P(T,h) &\propto e^{-f(T,h)+\tilde{f}(T,h)} .
\end{align}
From this we can write
\begin{align}
f(T,h)=\tilde{f}(T,h) -\log P(T,h) +\rm{const}.\label{eq:fcal}
\end{align}
Here, $P(T,h)$ can be obtained as the number of samples at each set of parameter values in a preliminary STM simulation. 
Thus, this equation enables one to refine the free energy much more easily than the reweighting methods, 
because the method does not require any iterations.
This method does not work well, however,  
when $P(T_i,h_j)$ is too small (or $\tilde{f}(T_i, h_j)$ is  too far away from true values) 
to obtain samples at $(T_i,h_j)$, while 
the reweighting techniques are still able to work. 
In the present work, we first used the reweighting methods to obtain rough estimates of the free energy for the entire parameter space.
We then used the combination of the reweighting methods and Eq. (\ref{eq:fcal}) for further refinements of the free energy.

Note that the WHAM gives another piece of information, namely DOS, which MBAR cannot directly calculate. However, 
the WHAM requires to make histograms before iterations and two kinds of calculations in an iteration step.
As the system size grows, the number of possible states increases. Thus, the calculation of DOS can be quite time-consuming. 
On the other hand, MBAR can be used without making histograms and one MBAR iteration step needs one equation. 
The length of one iteration, which is approximately proportional to the number of samples and parameter values, 
increases and can be time-consuming, as the system size is enlarged.
However, we have an impression that the MBAR is less time-consuming and more easily implemented than the WHAM. 
The parallelization of MBAR is slightly easier than that of WHAM and we actually did it with OPENMP.

\subsection{Temperature and external field distributions}
As is mentioned in the previous subsections, we have to give the set of temperature and external field values before ST or STM simulations.
Actually the determination involves trial and error. However, still the reweighting methods help one to do this
to a certain extent. 

Firstly the maximum and minimum values of temperature and external field were chosen so that the area of temperature and external field 
were wide enough to investigate the critical behaviors. This should be done separately for each system and what is to be investigated.

The distribution of temperature was chosen to be proportional to an exponential to the index number in small lattice sizes, 
as is common in simulated tempering and replica-exchange methods.  
However, in large lattice size systems, we assigned more number of values around $T_\mathrm{c}$ by hand.
More dense distribution was required where the heat capacity is large or the phase transition occurs.  
The distribution of external field is similarly assigned. In small lattice size it was proportional to the index of external field. 
However, in the larger lattice size, we assigned more points around $h=0$, in which the phase transition occurs. 
We assigned them in such a manner that the acceptance rate of ST parameter updates are preferably between 10\% and 50\%.
This fuzzy criterion is partly due to the two-dimensional distributions. 
A temperature distribution at a certain external field does not always give the same acceptance rates under another external field.

When the distributions of $T_i$ and $h_j$ turned out to be improper, we reassigned the distributions. 
In this case, we already had the samples and free energy estimates at a previous distribution, 
with which the reweighting method lets one to estimate the free energy at the newly distributed values.
Consequently, we did not have to start over the free energy calculations from the beginning. 
We actually repeated this parameter redistribution procedures several times, especially in large lattice size simulations.

\section{Results and Discussion}
\subsection{``Simulated Tempering and Magnetizing" (STM) simulations}
\begin{figure}[!hbtp]
   \begin{center}
       \includegraphics[width=5.5cm, clip , angle = 270]{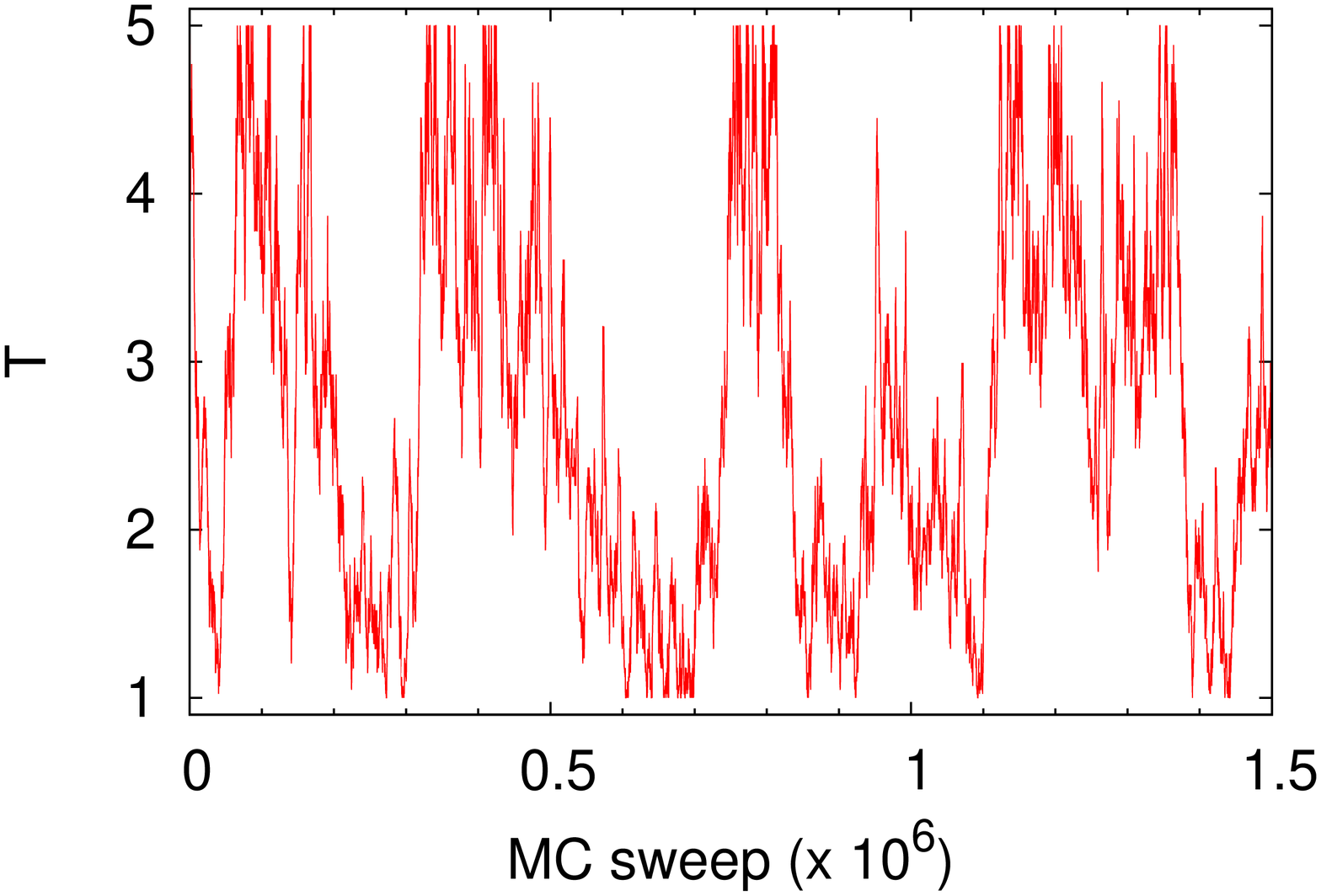}
       \caption{(Color online) The history of temperature $T$. The linear lattice size $L$ was 80.}
       \label{fig:temp}
   \end{center}
   \begin{center}
       \includegraphics[width=5.5cm, clip , angle = 270]{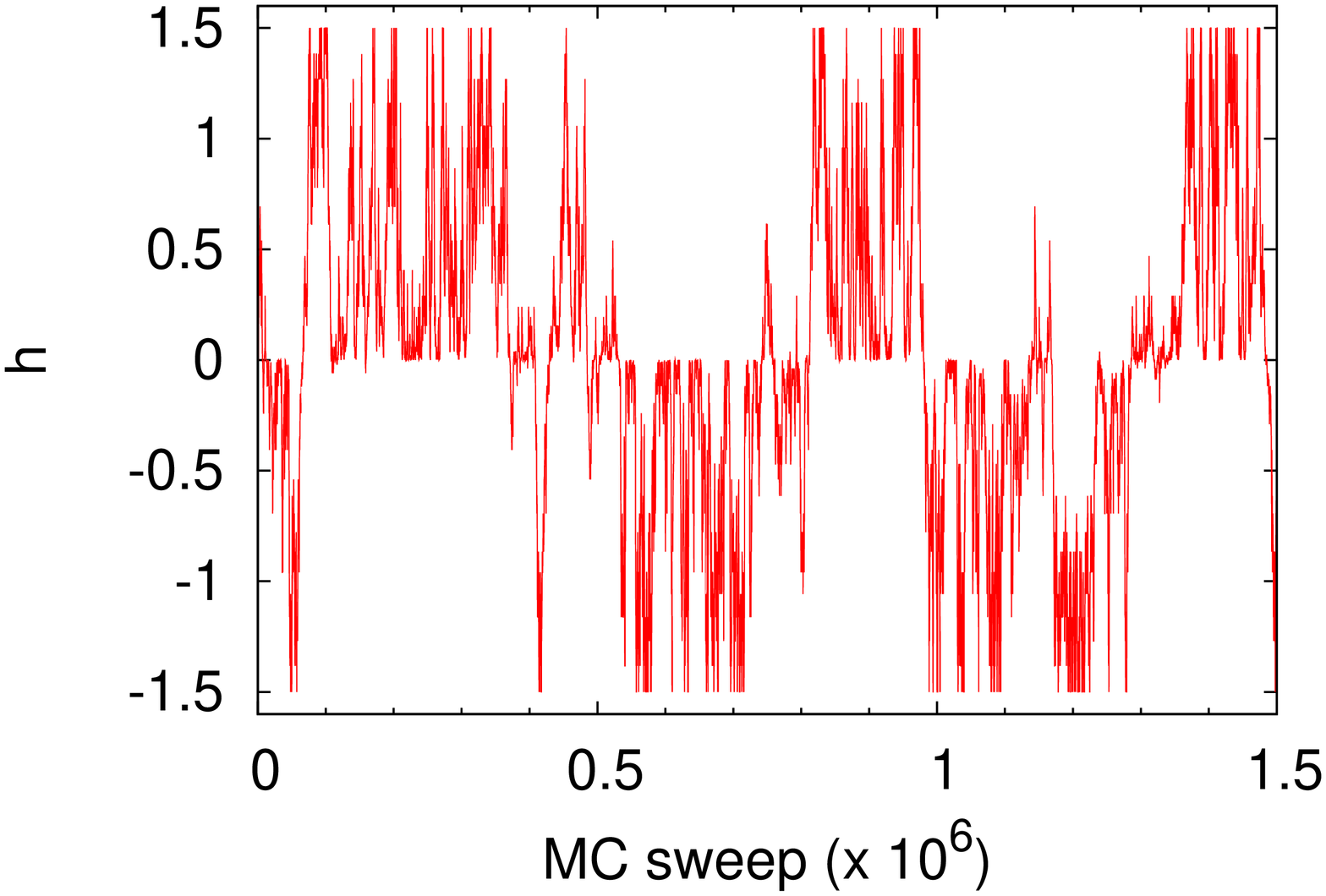}
       \caption{(Color online) The history of external field $h$. The linear lattice size $L$ was 80.}
       \label{fig:extf}
   \end{center}
\end{figure}

Firstly we shall  show that the two-dimensional ST simulations were carried out properly.
Figures \ref{fig:temp} and \ref{fig:extf} show temperature and external field, 
respectively, as functions of MC sweep.
Both were obtained from the simulations in which the linear lattice size was 80. 
The temperature and external field indeed realized random walks. 

\begin{figure}[!hbtp]
   \begin{center}
       \includegraphics[width=5.5cm, clip , angle = 270]{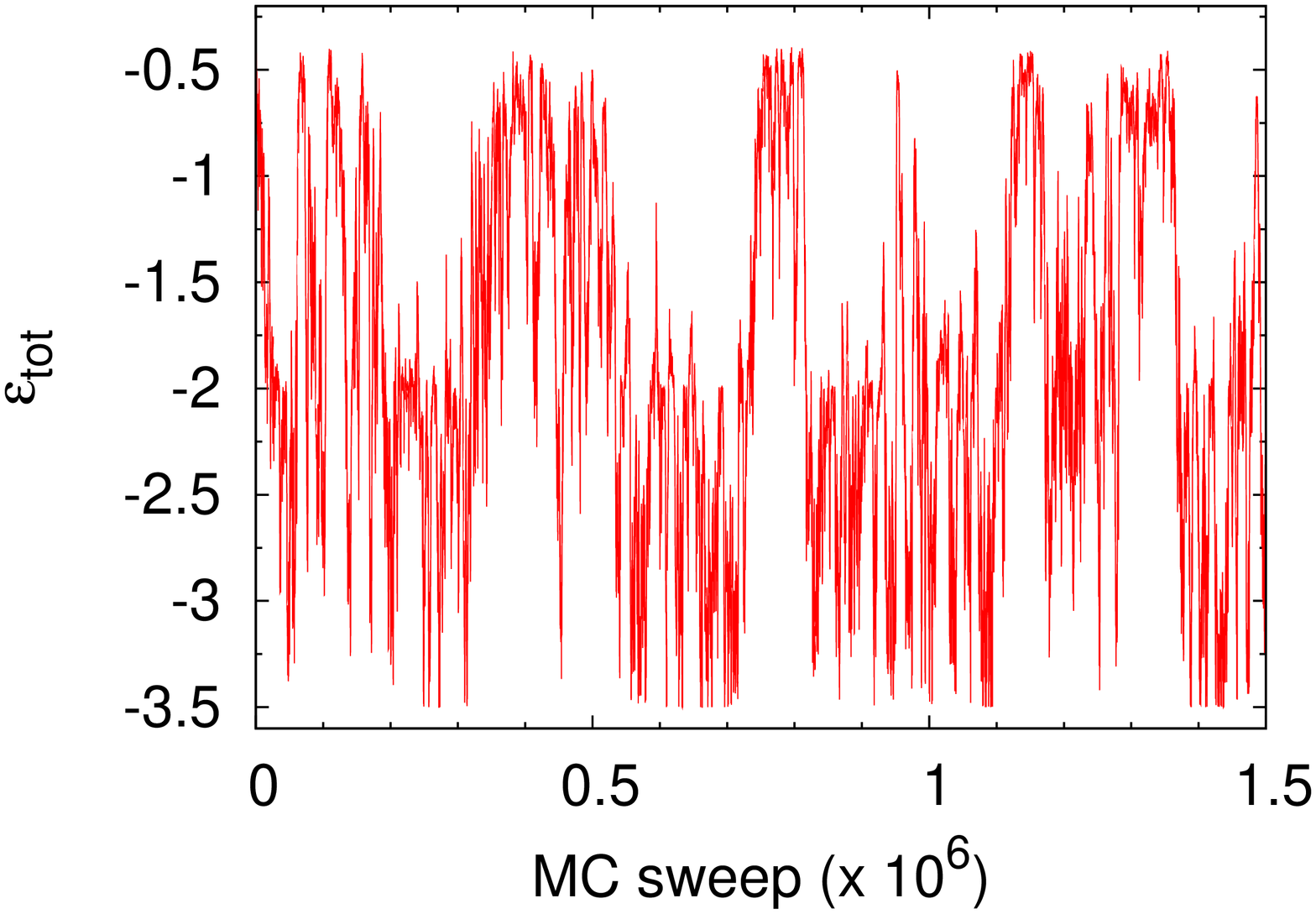}
       \caption{(Color online) The history of total energy per spin, $\epsilon_{tot}$. The linear lattice size $L$ was 80.}
       \label{fig:ener}
   \end{center}
   \begin{center}
       \includegraphics[width=5.5cm, clip , angle = 270]{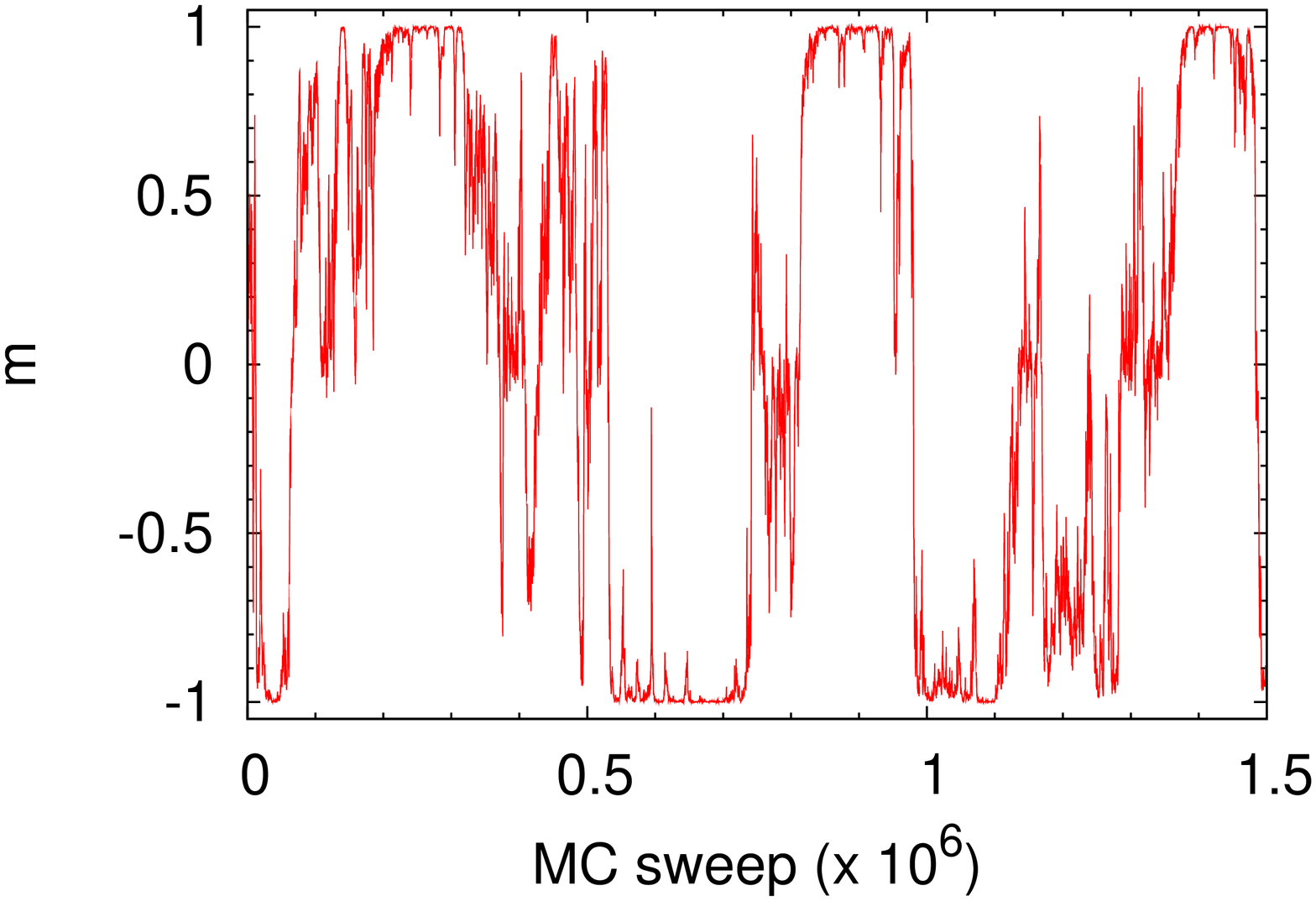}
       \caption{(Color online) The history of the magnetization per spin, $m$ ($\equiv M/L^2$). The linear lattice size $L$ was 80.}
       \label{fig:zika}
   \end{center}
\end{figure}

Figures \ref{fig:ener} and \ref{fig:zika} show energy and magnetization per spin, respectively, as functions of MC sweep. 
They also realized random walks. 
Note that there are  expected correlations between the temperature and energy (see Figs.\ \ref{fig:temp} and \ref{fig:ener})
and  between the external field and magnetization (see Figs.\ \ref{fig:extf} and \ref{fig:zika}).
The same behavior was observed in other lattice size simulations (data not shown).

\begin{figure}[hbtp]
    \begin{center}
        \includegraphics[width=7.5cm, clip ]{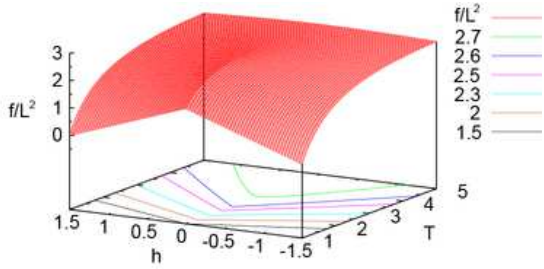}
        \caption{(Color online) The free energy per spin $f/L^2$ and 
                 its contour curves as a function of $T$ and $h$.The linear lattice size $L=80$.}
        \label{fig:fener}
    \end{center}
\end{figure}

Figure \ref{fig:fener} shows the dimensionless free energy per spin as a function of temperature and external field, 
which was obtained by applying MBAR to the results of the production runs.
Note that the partial differential of this free energy by $h$ gives $\frac{\left<m\right>}{T}$. 
The shape at $h=0$ suggests a jump of $m$  below $T_\mathrm{c}$, indicating existence of the first-order phase transitions.   

\begin{figure}[hbtp]
        \begin{center}
            \includegraphics[width=7.5cm, clip ]{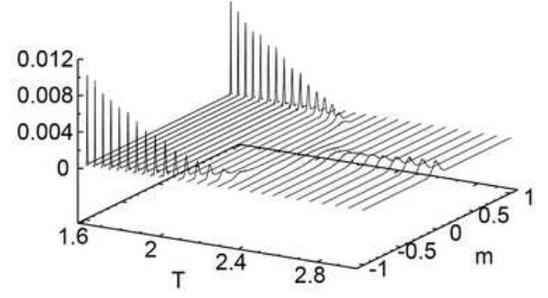}
            \caption{ The distribution of $m$ as a function of $T$ for $h=0$. The linear lattice size $L$ was 80. }
            \label{fig:MdistT}
        \end{center}
\end{figure}
\begin{figure}[hbtp]
            \includegraphics[width=7.5cm, clip ]{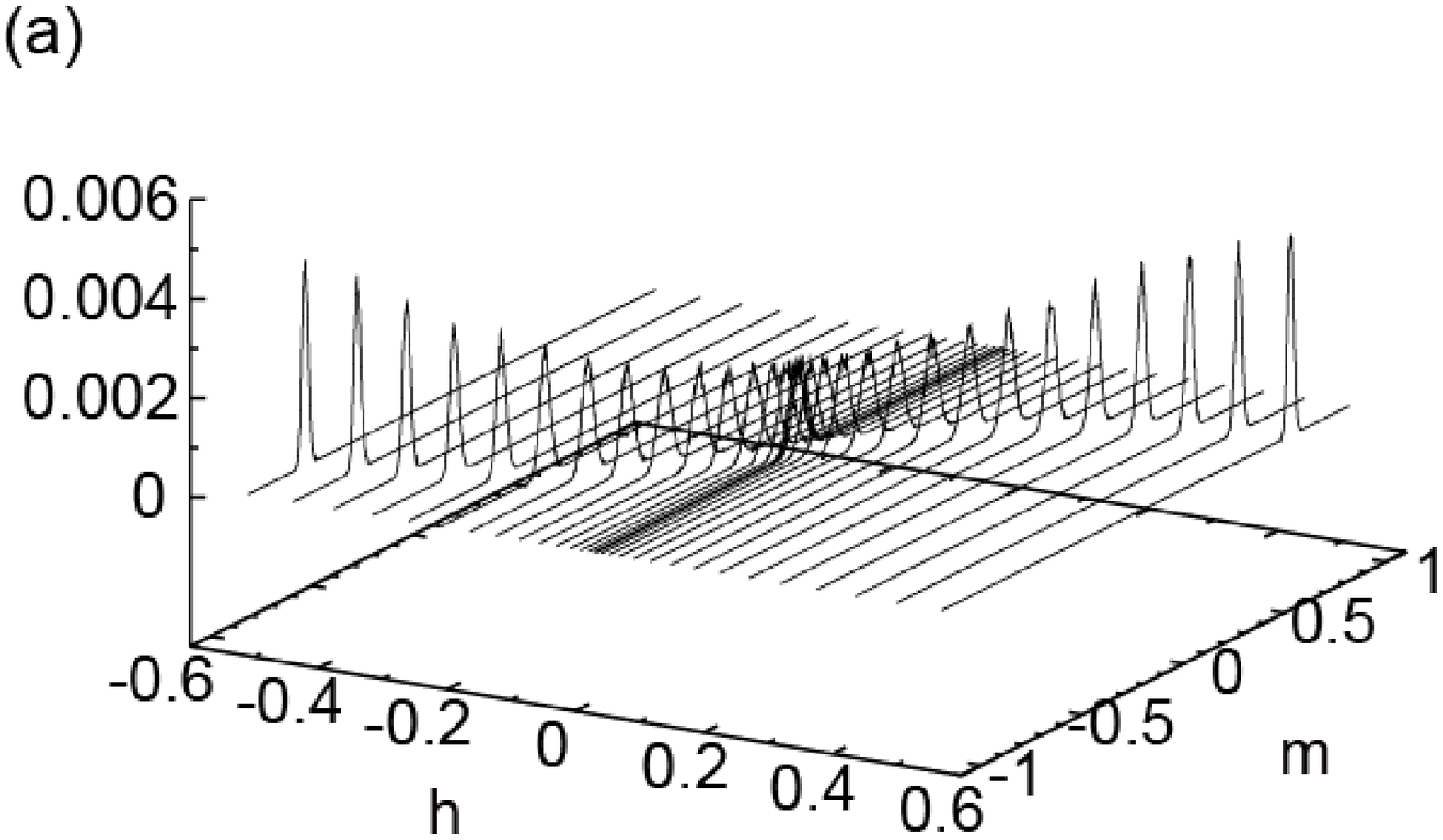}
            \includegraphics[width=7.5cm, clip ]{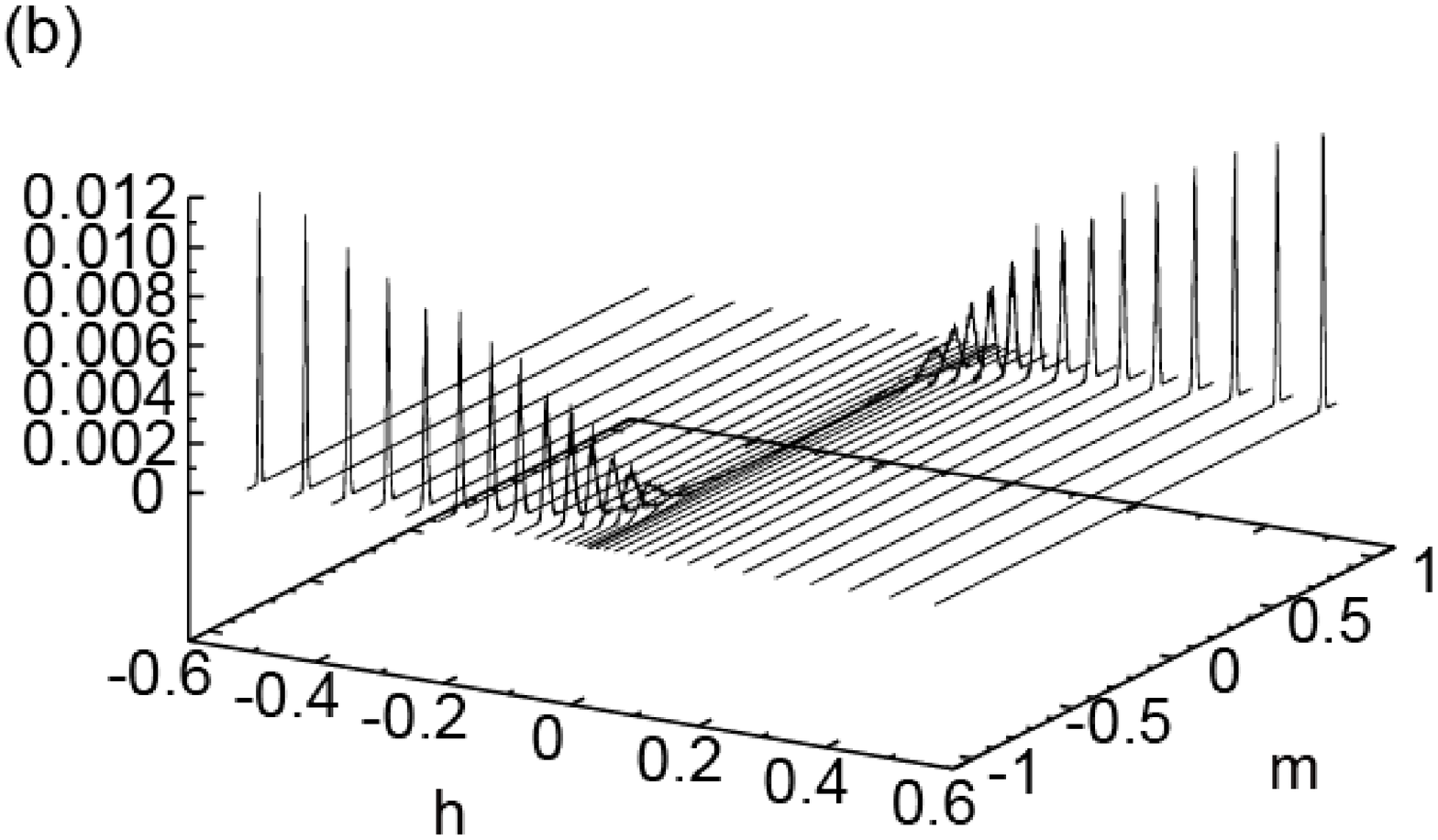} \\
            \includegraphics[width=7.5cm, clip ]{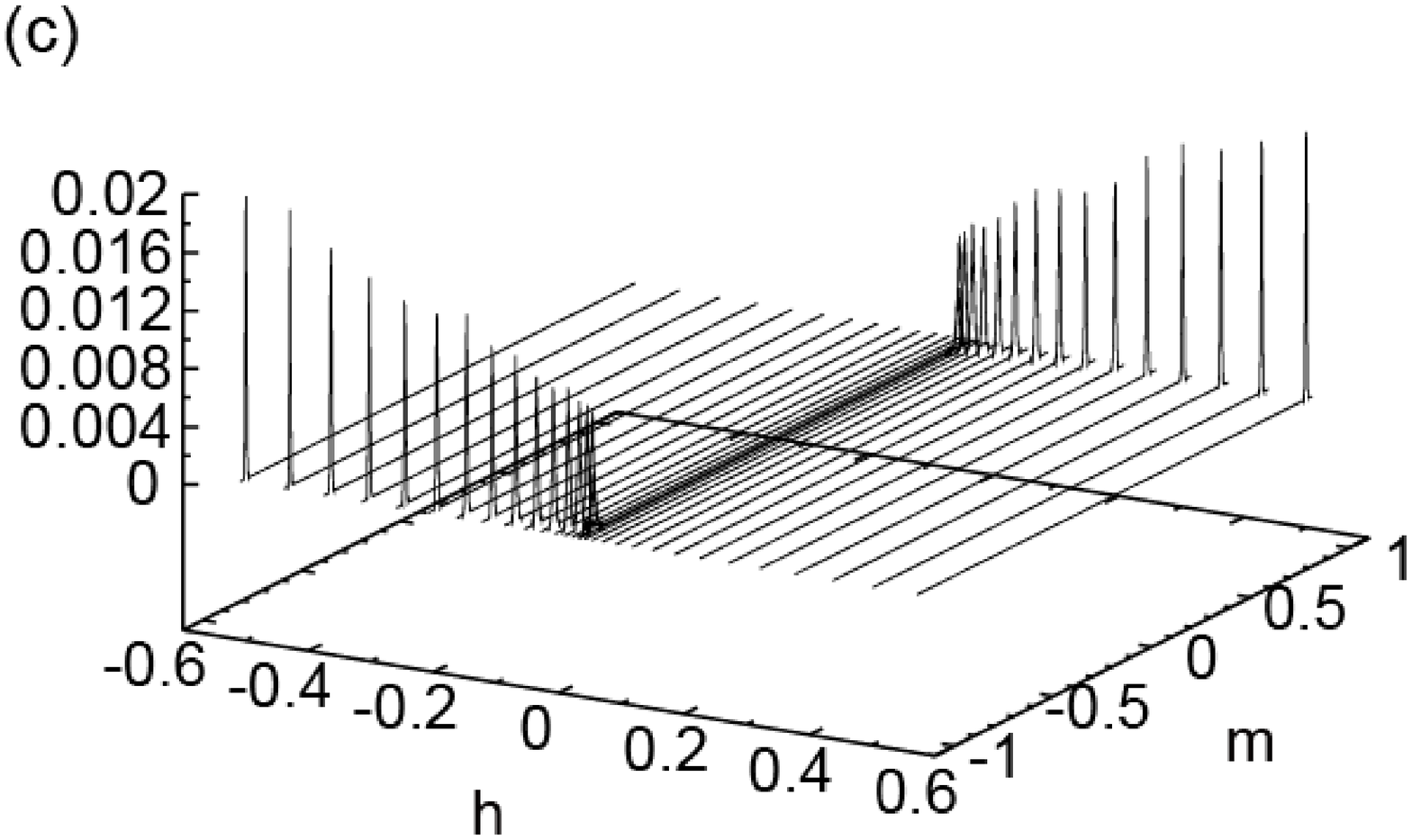}
            \caption{ The distribution of $m$ as a function of $h$ when 
                      (a) $T=3.21$, (b) $T=2.316$ and (c) $T=1.967$. The linear lattice size $L$ was 80.}
            \label{fig:MdistH}
\end{figure}

Figure \ref{fig:MdistT} shows the distribution of magnetization as a function of temperature.
Below $T_\mathrm{c}$ the distribution is separated into two parts. 
As temperature increases, the distribution becomes broader. Near $T_\mathrm{c}$ 
the distribution is the broadest and two peaks merge. It then becomes narrower. 
Note that this figure was obtained by only four production runs (see Table \ref{table}), 
and can be obtained even by only one production run, though the error is expected to become larger. 
Figures \ref{fig:MdistH}(a), \ref{fig:MdistH}(b), and \ref{fig:MdistH}(c) show the distribution of magnetization 
as a function of external field above, around, and below $T_\mathrm{c}$, respectively.
Above $T_\mathrm{c}$, the change is smooth and continuous (see Fig.\ \ref{fig:MdistH}(a)). 
Around $T_\mathrm{c}$, the distribution becomes very wide around $h=0$  (see Fig.\ \ref{fig:MdistH}(b)).
This is one of the properties of the second-order phase  transitions. 
Below $T_\mathrm{c}$, the distribution jumps from one side to the other side at $h=0$ (see Fig.\ \ref{fig:MdistH}(c)). 
This abrupt jump of distribution is one of the properties of the first-order phase transitions.

\begin{figure}[hbtp]
        \begin{center}
            \includegraphics[width=5.5cm, clip, angle=270 ]{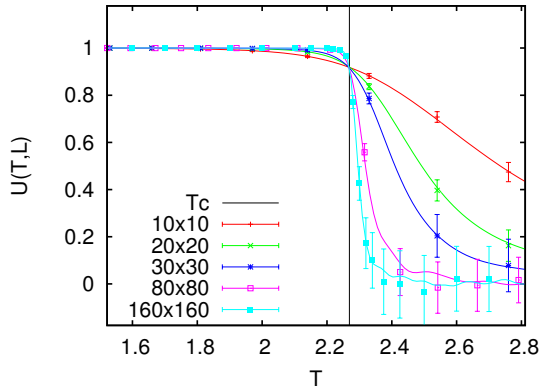}
            \caption{(Color online) Binder cumulant $U$ vs temperature.}
            \label{fig:Binder1}
        \end{center}
\end{figure}

We also calculated the Binder cumulant \cite{binder1981finite} defined by 
\begin{align}
U(T,h,L) \equiv \frac{1}{2}\left( 3-\frac{\left<m^4 \right>}{\left<m^2 \right>^2}\right) .
\end{align}
Figure \ref{fig:Binder1} shows the Binder cumulant as a function of temperature. 
As is well-known, the graphs cross at one point at $T_\mathrm{c}$. 
The error bars were obtained by the jackknife method \cite{Miller1974,berg2004book}.  

\begin{figure}[hbtp]
        \begin{center}
            \includegraphics[width=5.5cm, clip, angle=270 ]{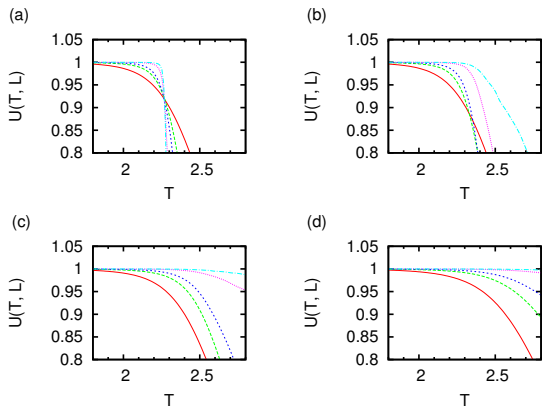}
            \caption{(Color online) Binder cumulant $U$ vs temperature under different external fields. 
                     (a) $h=0$. (b) $h=0.01$. (c) $h=0.05$. (d) $h=0.1$. The dotted green curve, solid red curve,
                     and dashed blue curve stand for the results for $L=20$, $80$, and $160$, respectively.}
            \label{fig:Binder2}
        \end{center}
\end{figure}

Figure \ref{fig:Binder2} shows the Binder cumulant as a function of temperature under different external fields. 
The graphs do not cross at one point in the presence of finite external field. The amount of errors is expected to be 
on the same level of Fig.\ \ref{fig:Binder1} and the error bars are suppressed here to aid the eye.   

\subsection{Comparison of ST with STM}
\begin{figure}[hbtp]
        \begin{center}
            \includegraphics[height=5.5cm, clip]{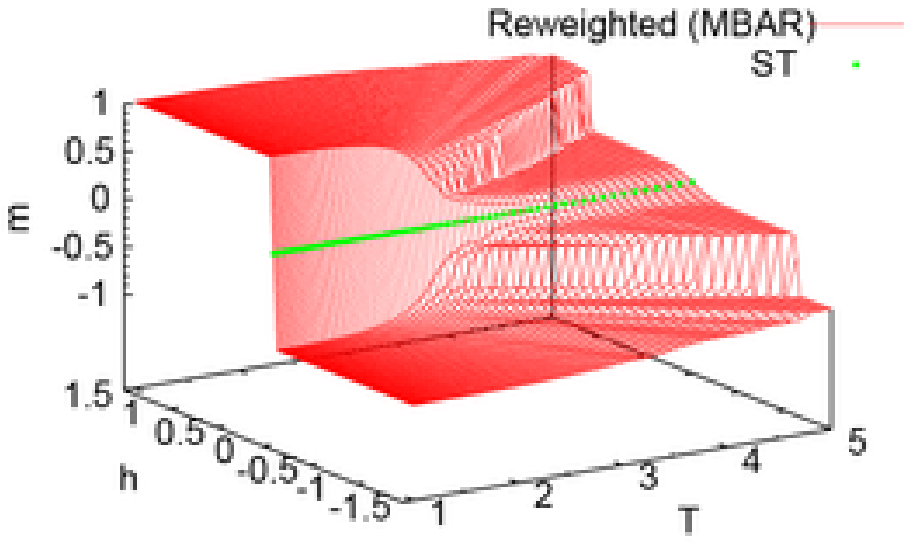}
            \caption{(Color online) Reweighted data (red) and original data (green) obtained by the conventional ST.
                      The linear lattice size $L$ was 80. }
            \label{fig:mag_rew}
        \end{center}
        \begin{center}
            \includegraphics[height=5.5cm, clip]{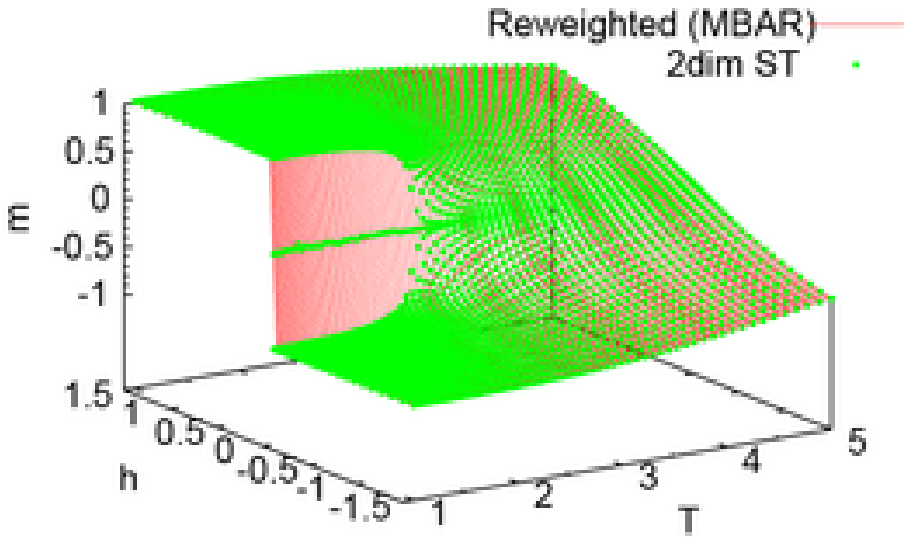}
            \caption{(Color online) Reweighted data (red) and original data (green) obtained by the STM simulations.
                      The linear lattice size $L$ was 80. }
            \label{fig:mag_rew_2dim}
        \end{center}
        \begin{center}
            \includegraphics[width=7.5cm]{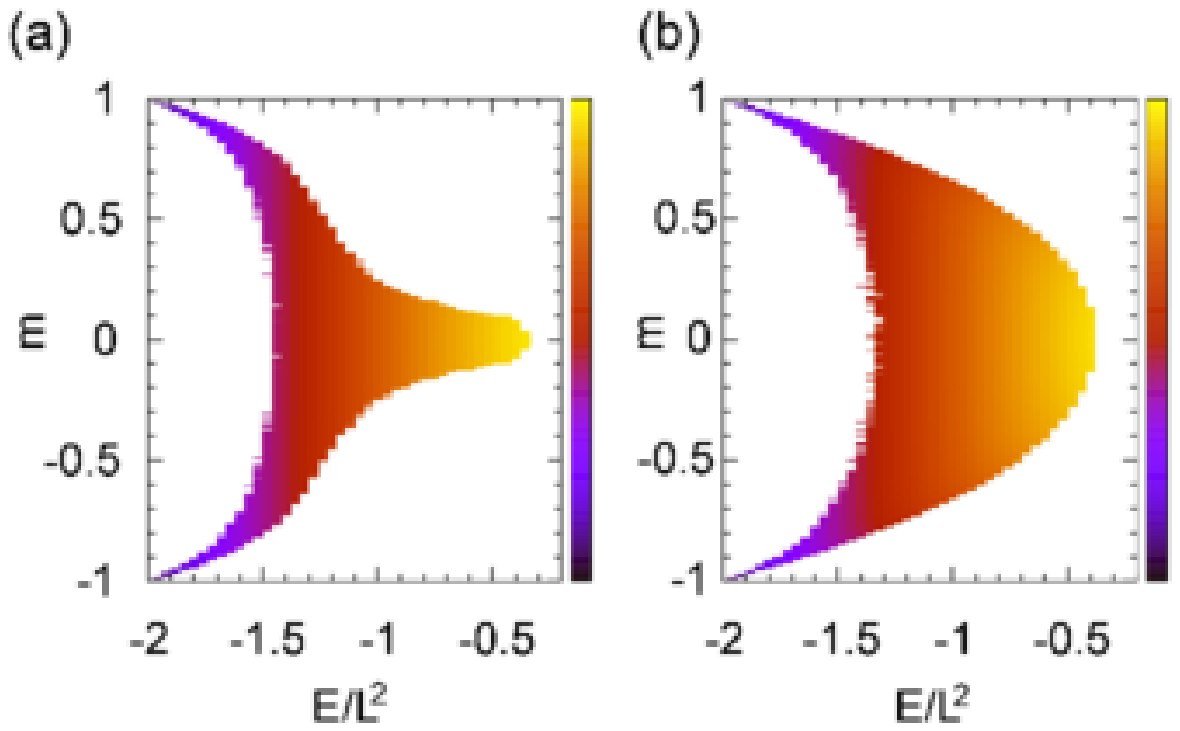}
            \caption{(Color online) Calculated DOS obtained by WHAM with ST (a) and STM (b) data, respectively. 
                      The linear lattice size $L$ was 80.}
            \label{fig:dos}
        \end{center}
\end{figure}

We compared the results of the STM method with those of the conventional ST method.  
Figures \ref{fig:mag_rew} and \ref{fig:mag_rew_2dim} show the magnetization as a function of temperature and external field, 
which was calculated using MBAR with the data obtained by the conventional ST and STM simulation, respectively.   
Figure \ref{fig:mag_rew} obtained by the conventional ST shows artifact jumps 
at a high temperature and a certain external field. 
This must have been caused by a failure of sampling  some parts of states.
On the other hand, the results by the STM simulations are smooth (see Fig.\ \ref{fig:mag_rew_2dim}).
Figure \ref{fig:dos} shows the density of states obtained by conventional ST and STM simulations. 
This obviously illustrates that the area in which the energy is relatively high with somewhat strong magnetizations 
were not sampled by the conventional ST method. 
These results imply that the dimensional extension in the STM enlarged the sampled space.

\subsection{Simulated Magnetizing (SM)}
\begin{figure}[hbtp]
     \begin{center}
         \includegraphics[width=5.cm, clip , angle = 270]{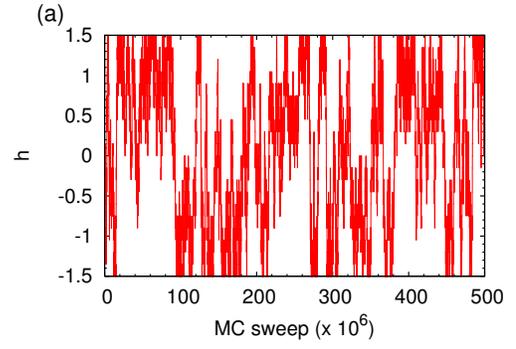}
         \includegraphics[width=5.cm, clip , angle = 270]{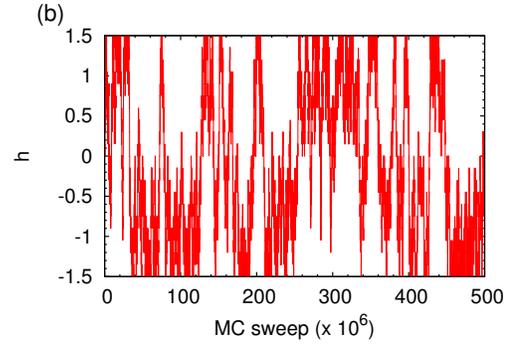}
         \includegraphics[width=5.cm, clip , angle = 270]{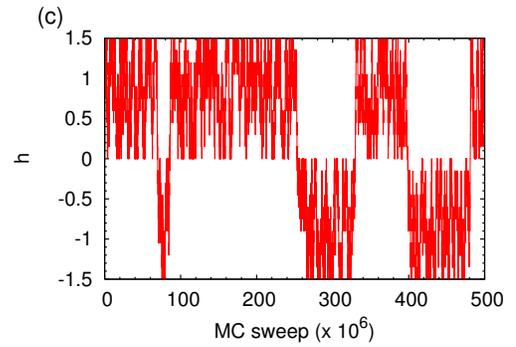}
         \includegraphics[width=5.cm, clip , angle = 270]{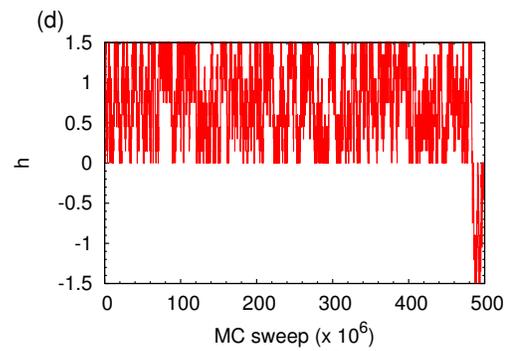}
         \caption{(Color online) External field vs MC sweep in SM simulations under $T_\mathrm{c}$ ($T=1.97$). 
                  The linear lattice size $L$ is (a) $2$, (b) $4$, (c) $8$, and (d) $10$, respectively.}
         \label{fig:sm_underTc}
     \end{center}
\end{figure}

\begin{figure}[hbtp]
     \begin{center}
         \includegraphics[width=5.cm, clip , angle = 270]{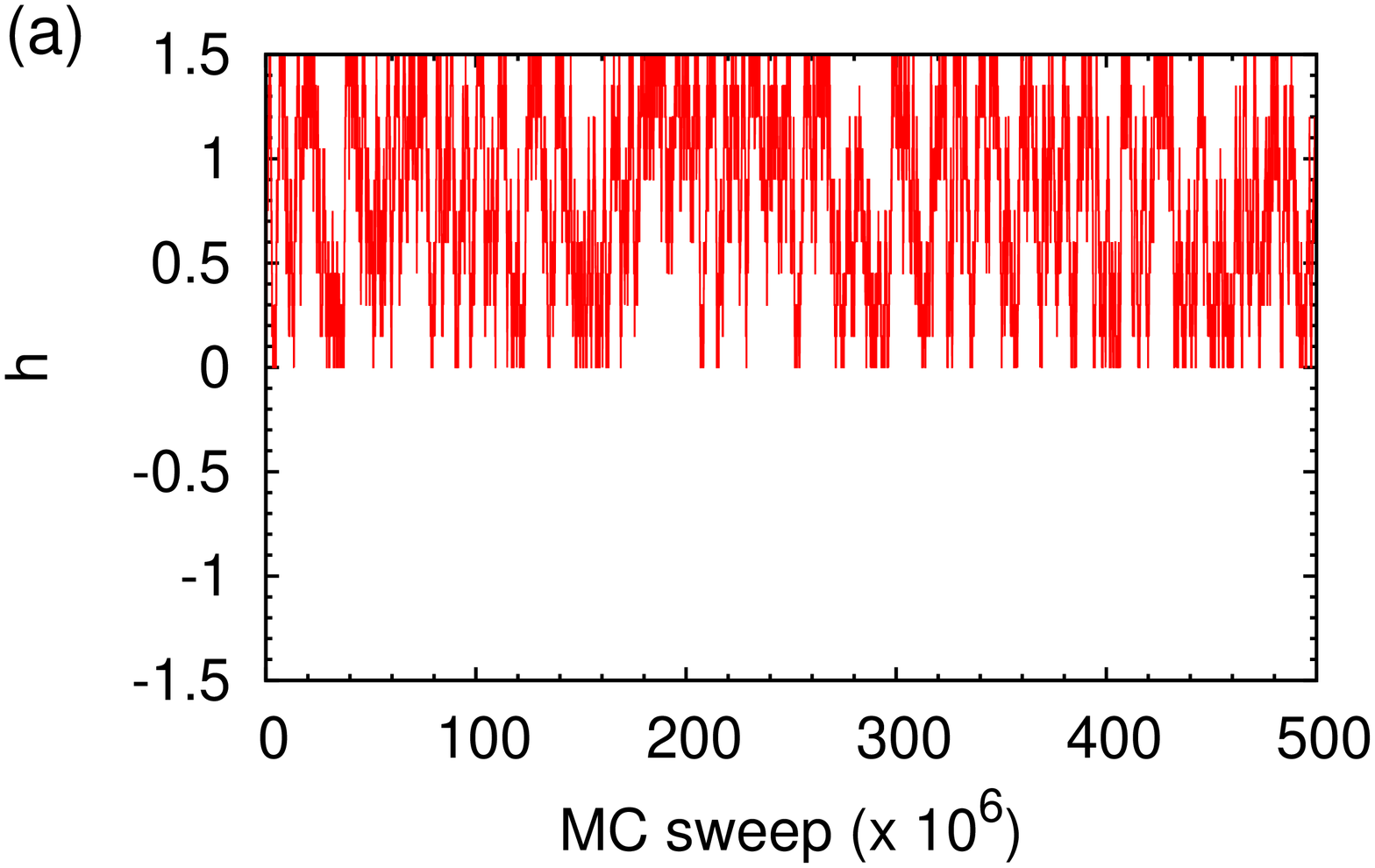}
         \includegraphics[width=5.cm, clip , angle = 270]{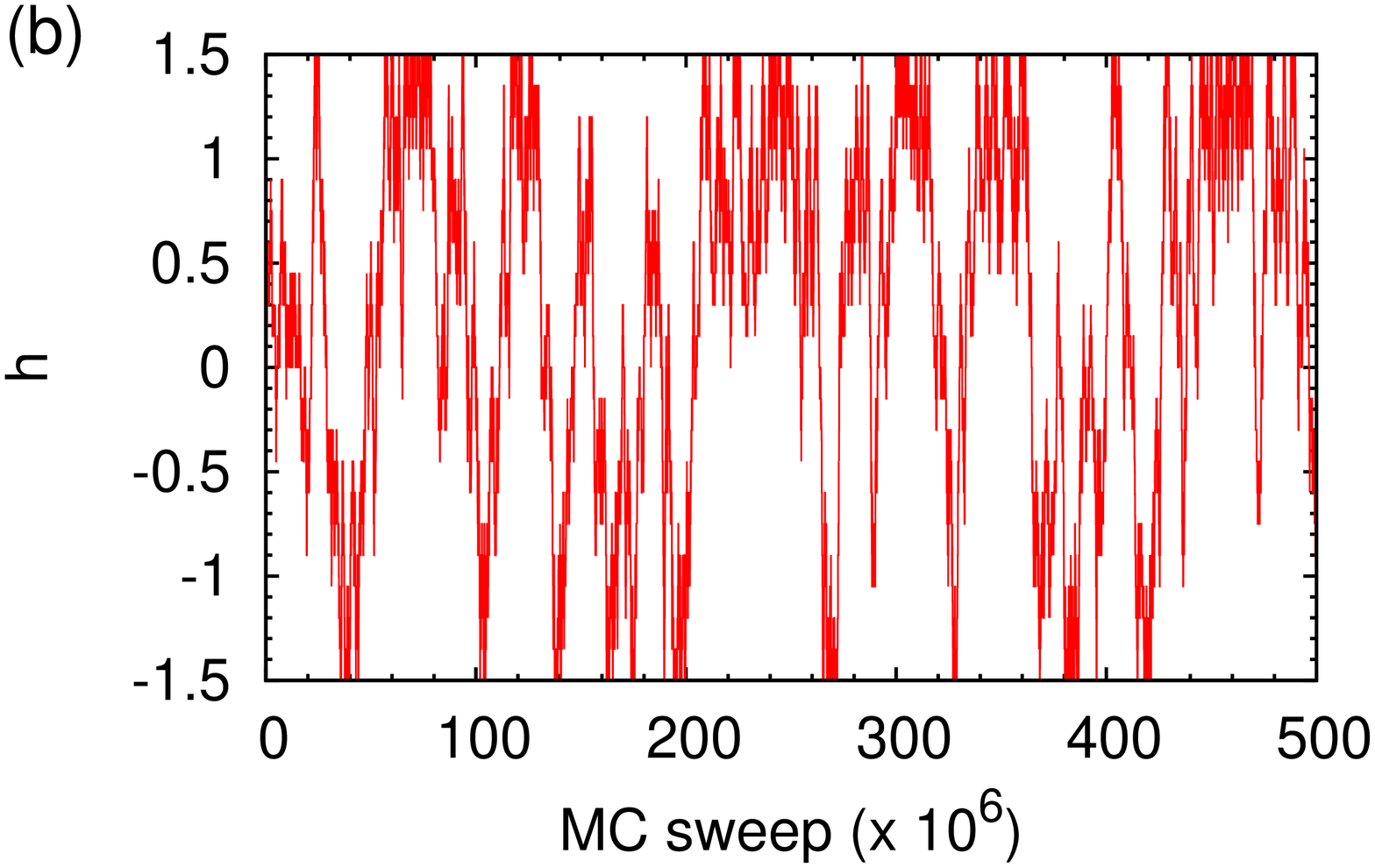}
         \caption{(Color online) External field and MC sweep in the SM simulation 
                  (a) under $T_\mathrm{c}$ ($T=1.97$) and (b) above $T_\mathrm{c}$ ($T=3.88$). The linear lattice size $L$ is 20. }
         \label{fig:sm_comp}
     \end{center}
\end{figure}

We study the compatibility of ST with the first-order phase transition along external field changes, 
by performing ``Simulated Magnetizing" (SM) simulations, 
in which the temperature is fixed and the external field is updated by the Metropolis criteria.
Figure \ref{fig:sm_underTc} shows the external field as a function of MC sweep in the SM simulations below $T_\mathrm{c}$.
We performed SM simulations in a number of lattice sizes from 2$\times$2 to 20$\times$20. 
These graphs illustrate the fact that as the system size becomes larger, the difficulty in simulations grows. 
In fact it finally became impossible to observe the events in which the magnetization goes to 
the other side across the zero point (see Fig.\ \ref{fig:sm_comp}(a)),
while it was still possible above $T_\mathrm{c}$ (see Fig.\ \ref{fig:sm_comp}(b)).
These results imply that the full range random walk happens above $T_\mathrm{c}$ but not below $T_\mathrm{c}$.
Therefore, this result suggests that the random walk of temperature is crucial for the full range random walk of external field.
The full range random walk of the external field happens 
in the STM  simulation when the temperature was high above $T_\mathrm{c}$.
Note that the Ising model is equivalent to the lattice gas model \cite{lee1952statistical}. 
Hence, what happened in STM simulations can be understood as that 
even though the phase transitions between gas and liquid do not directly occur, 
they do occur through the ``super critical water region."

  
To explore this phenomenon more clearly, we present Video  \ref{video}, 
which shows how the temperature and external field changed during the STM simulations. 
The picture shown is the first frame of the video. The red line is drawn where the first-order phase transitions occur.
The movement of parameter can touch the red line but rarely go across the red line below $T_\mathrm{c}$. 
On the contrary, above $T_\mathrm{c}$, the external
field can change smoothly.       

\begin{video}
    \includegraphics[width=5.cm, clip , angle = 270]{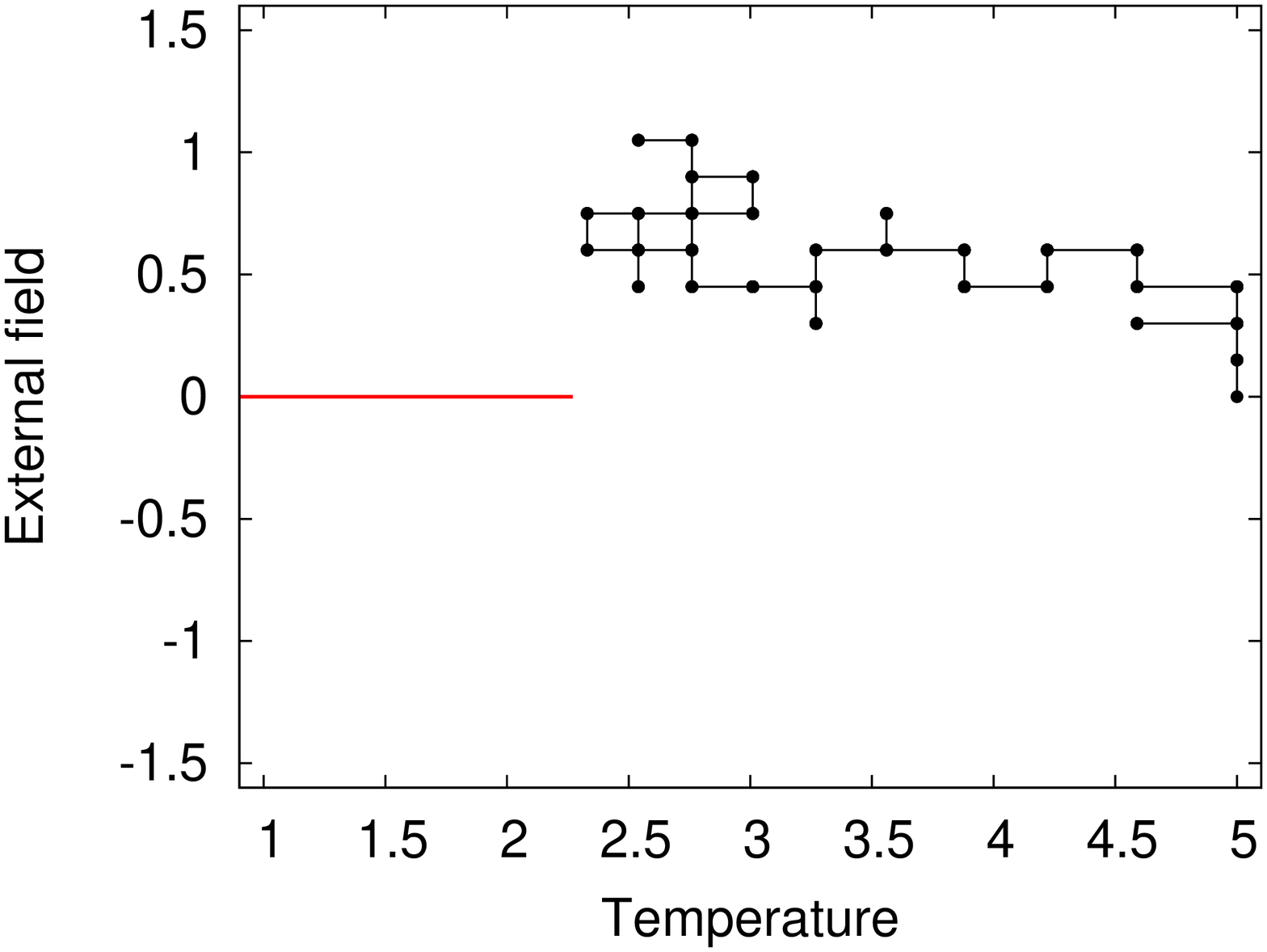}
    \\
    \setfloatlink{http://www.tb.phys.nagoya-u.ac.jp/\%7etnagai/published/guide.html}
    \caption{\label{video} (Color online)  
             The red horizontal line is drawn between $T=0$ and $2.27$ ($\approx  T_\mathrm{c}$) at $h=0$.
             Black points show the sampled set of values of temperature and external field. 
             Black lines are drawn when the update is accepted between the conditions. 
             Parameters within 5000 MC sweeps are shown, during which 100 parameter-updating attempts were made.
             The windows change every 1000 MC sweeps. The linear lattice size $L$ was 20.
             }
\end{video}

\subsection{How often temperature or external field should be updated?}
A common question about this kind of simulation is how frequently the parameter-updating attempts should be made.
We want to emphasize  that as long as the detailed balance condition is satisfied the simulations should be 
correctly carried out.  

\begin{figure}[hbtp]
        \begin{center}
            \includegraphics[width=5.5cm, clip, angle = 270 ]{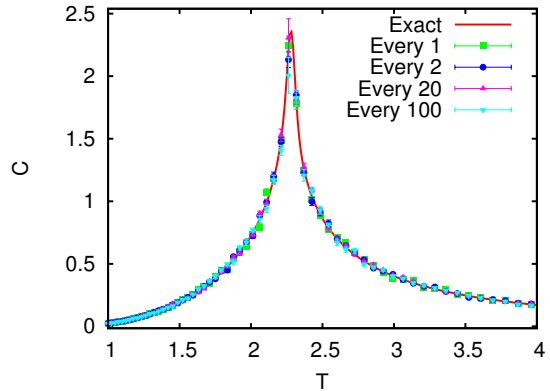}
            \caption{(Color online) Heat capacity per spin, $C$, at $h=0$. 
                        The linear lattice size $L$ was 80.  As the legends shown in the figure, green square,
                        blue circle, purple triangle, and light-blue inverse-triangle represent 
                        that one parameter-updating attempt is made every one, two, twenty, and a hundred MC
                        sweeps, respectively. The exact result (red solid line) was obtained 
                        by Berg's program \cite{berg2004book} based on Ref.\ \cite{Ferdinand1969IsingFinite}.}
            \label{fig:c_t_STfrq}
        \end{center}
\end{figure}

We compared STM simulations performed with different parameter-updating frequencies.
Figure \ref{fig:c_t_STfrq} shows the results of the heat capacity as a function of temperature at $h=0$, 
which were obtained by the STM method with different conditions. 
The conditions are one parameter-updating attempt every one, two, twenty,
and a hundred MC sweeps.  
They show good agreement with each other. The error bars were obtained by the jackknife method \cite{Miller1974,berg2004book}. 
Note that the error bars tend to be larger as the parameter-updating frequency becomes less. 

\begin{figure}[hbtp]
        \begin{center}
            \includegraphics[width=5.5cm, clip, angle = 270 ]{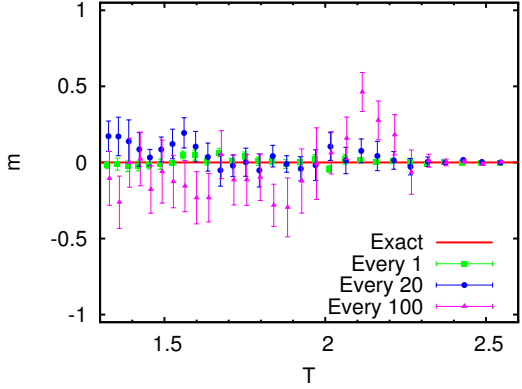}
            \caption{(Color online) Magnetization per spin $m$ when $h=0$. 
                      As the legends shown in the picture, the green square, blue circle and purple triangle
                    represent that one parameter-updating attempt is made every one, twenty, and a hundred MC sweeps, respectively. 
                    Some error bars were slightly shifted horizontally to aid the eye. }
            \label{fig:M_t_STfrq}
        \end{center}
\end{figure}

Figure \ref{fig:M_t_STfrq} shows the magnetization as a function of temperature at $h=0$. 
Data were obtained with several parameter-updating frequencies, 
such as one parameter-updating attempt every one, twenty, and a hundred MC sweeps.
They  also  agree with each other. 
Note that because finite sizes are employed, the magnetization under $T_\mathrm{c}$ at $h=0$ is also zero. 
With the lower parameter-updating frequency, the convergence was not so good and the error bars tend to be larger. 
The error bars were obtained by the jackknife method \cite{Miller1974,berg2004book}.
These results suggest that the frequent parameter update does not make any artifacts and that it should be recommended.         

\begin{figure}[hbtp]
        \begin{center}
            \includegraphics[width=8.5cm, clip ]{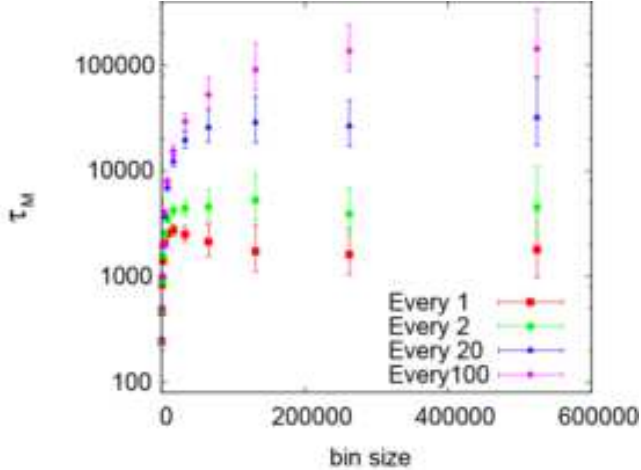}
            \caption{(Color online) Correlation time analysis. Error bars show the 95\% confident interval.  }
            \label{fig:cor_time}
        \end{center}
\end{figure}

Figure \ref{fig:cor_time} shows the integrated correlation time of magnetization obtained at different parameter-updating frequencies. 
The height of data is expected to converge to the integrated correlation time between samples. 
This was calculated by using the jackknife method with different bin sizes \cite{Miller1974,berg2004book}.  
Data were stored every ten MC sweeps.
Thus, the correlation time measured by MC sweep should be  ten times larger.
The error bars were obtained with the $\chi ^2$ distribution.
These results suggest that the higher parameter-updating frequency was employed, the lower 
correlation time was obtained. Therefore, frequent parameter updates are preferred.
Note that the observation that the frequent parameter updates are preferable 
is in accord with the statement that frequent replica-exchanging attempts 
are recommended  \cite{sindhikara2008exchange,sindhikara2010}.

\subsection{Observation of crossover}
We study the crossover behavior of the phase transitions. 
We calculated the magnetization by MBAR around the critical point. 

We employ the finite-size scaling approach, which is discussed in Ref. \cite{PhysRevB.13.2997}.
The scaling form of magnetization $m$ with respect to temperature and external field is given by
\begin{align}
mL^{\beta/\nu}&= \Psi (L^{1/\nu}t,L^{(\gamma +\beta)/\nu}h) \,,
\end{align}
where $t=|T-T_\mathrm{c}|/T_\mathrm{c}$  and $L$ is the linear size of lattice. 
The Greek letters $\nu$ and $\gamma$ stand for critical exponents.
In the two-dimensional Ising model, $\beta = 1/8$, $\delta = 15$, $\nu =1 $, and $\gamma = 7/4$.  

\begin{figure}[hbtp]
        \begin{center}
            \includegraphics[width=5.5cm, clip, angle = 270 ]{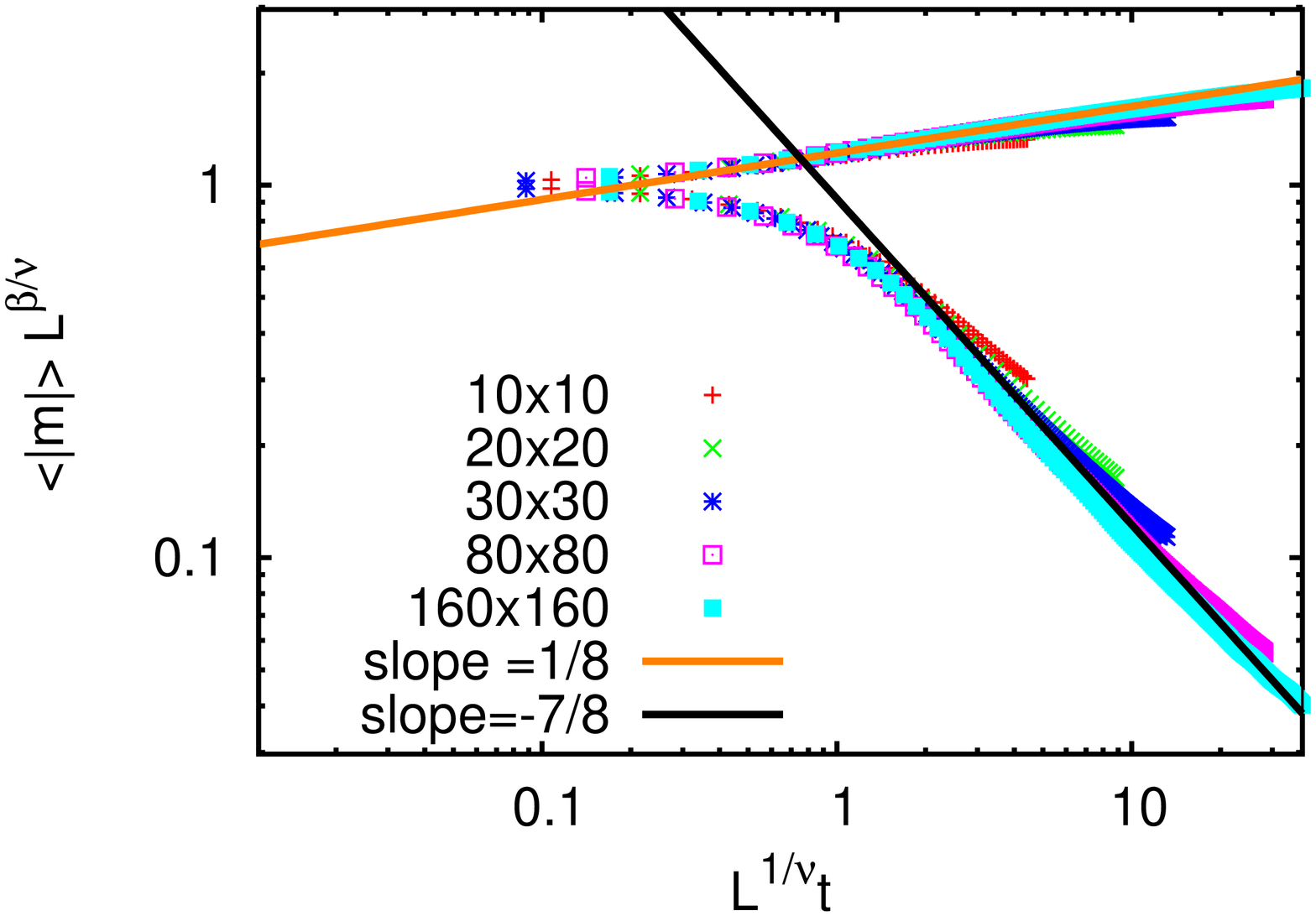}
            \caption{(Color online) Scaled $m$ when $h=0$. The lines are the same as used in Ref.\ \cite{PhysRevB.13.2997}. }
            \label{fig:scaledM}
            \includegraphics[width=5.5cm, clip ,angle = 270 ]{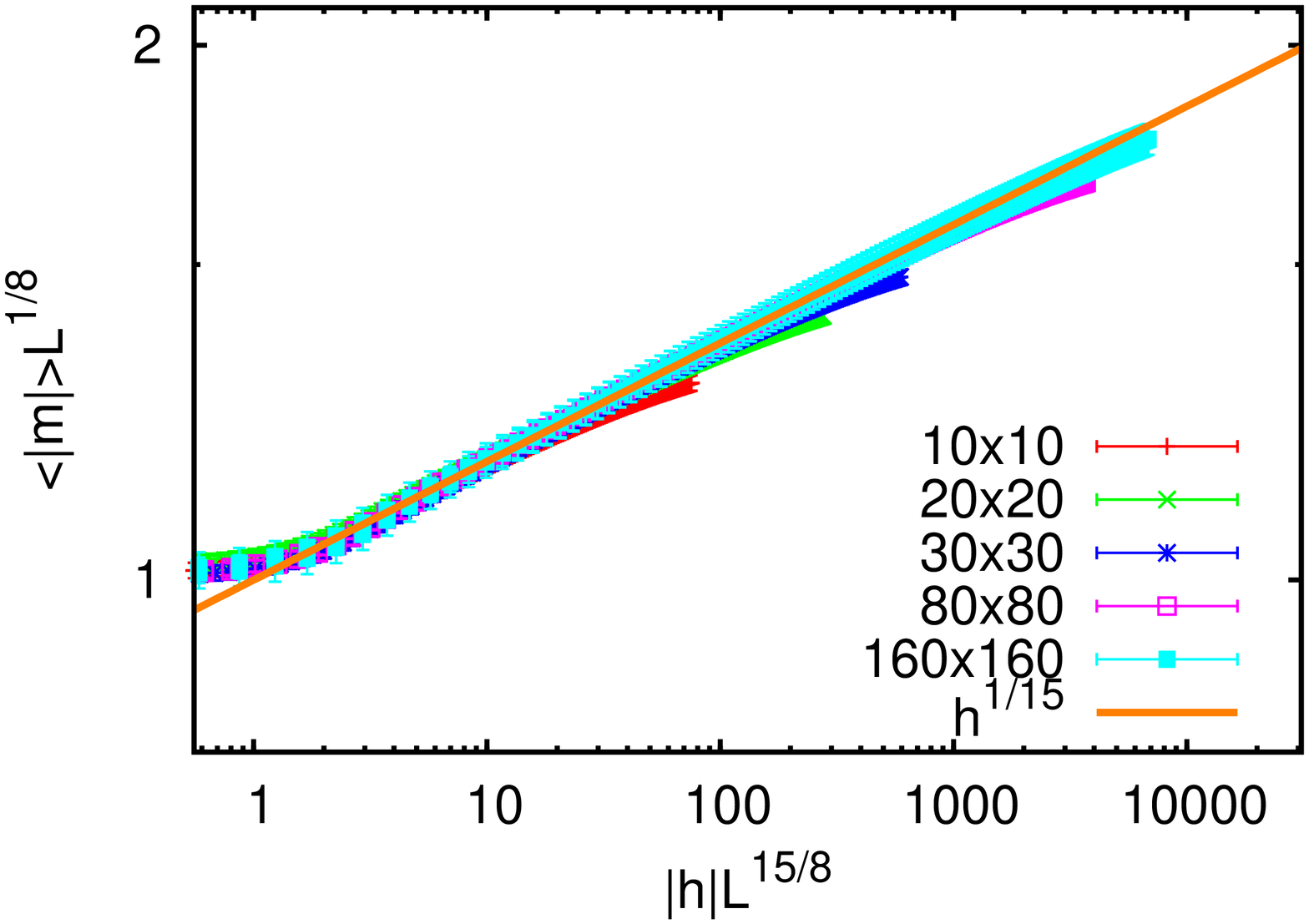}
            \caption{(Color online) Scaled $m$ at $T=T_\mathrm{c}$. }
            \label{fig:scaledM2}
        \end{center}
\end{figure}

Firstly we examine  the scaling behavior of the magnetization.
Figures \ref{fig:scaledM} and \ref{fig:scaledM2} show the magnetization as functions of $T$ and $h$, respectively, 
and we see that it obeys the critical behavior of 
$m \sim |T-T_\mathrm{c}|^{\beta}$ and $m \sim |h|^{1/\delta}$, respectively. 
According to the scaling approach, when $Lt$ or $L^{15/8}h$ is 
large enough, then the finite effect can be negligible. 
In this case, Figs.\ \ref{fig:scaledM} and \ref{fig:scaledM2} imply that 
those conditions are given by $Lt>0.2$ and $L^{15/8}h>1.1$, respectively. 

\begin{figure}[hbtp]
        \begin{center}
            \includegraphics[width=5.5cm, clip , angle = 270 ]{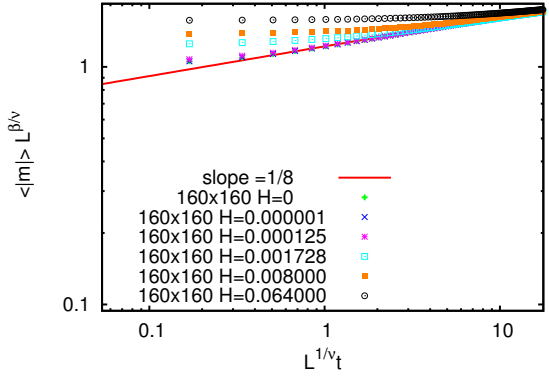}
            \caption{(Color online) Scaled $m$ near $h=0$. }
            \label{fig:scaledMwithH}
        \end{center}
\end{figure}

We  now  study the behavior under the conditions slightly different from the critical point.
Figure \ref{fig:scaledMwithH} shows the magnetization as a function of temperature near $h=0$. 
As the external field was increased, the 
behavior was differentiated in the low temperature region. 
Even in the presence of weak external field, 
the magnetization obeys $t^{1/8}$ when the temperature was relatively high enough. 
However, with relatively strong external field, the scaling behavior disappears.   

\begin{figure}[hbtp]
     \begin{center}
         \includegraphics[width=5.5cm, clip, angle = 270]{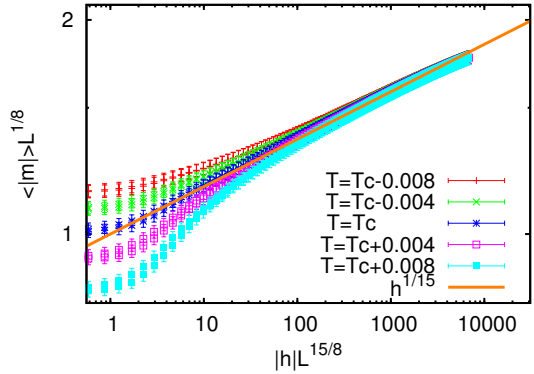}
         \caption{(Color online) Scaled $m$ near $T=T_\mathrm{c}$. }
         \label{fig:scaledM2withT}
     \end{center}
\end{figure}

Figure \ref{fig:scaledM2withT} shows the magnetization as a function of external field 
near $T=T_\mathrm{c}$. As the temperature was deviated from $T_\mathrm{c}$, the 
behavior is differentiated in the weak external field region. 
Thus, even with slight difference from $T_\mathrm{c}$, the magnetization obeys $h^{1/15}$ 
when the external field is strong enough. 

\begin{figure}[hbtp]
        \begin{center}
            \includegraphics[width=5.5cm, clip, angle = 270]{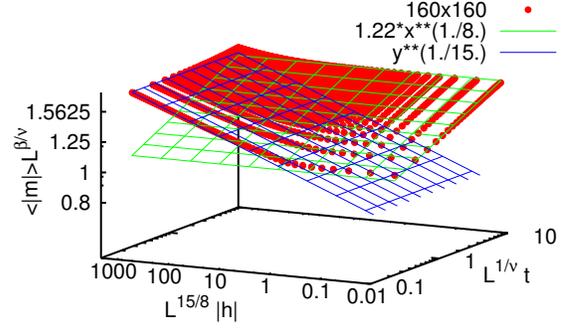}
            \caption{(Color online) Scaled $m$ about the critical point. The linear lattice size $L$ was 160. 
                     We only display the results for $T<T_\mathrm{c}$.   }
            \label{fig:3d_scaledM}
        \end{center}
\end{figure}

Figure \ref{fig:3d_scaledM} illustrates the comprehensive behavior of $\left<|m|\right>$ near the critical point. 
Note that this is a log scale plot. Near the $h$-axis $\left<|m|\right>$ obeys $|h|^{1/15}$ and near the $T$-axis
$\left<|m|\right>$ obeys $|t|^{1/8}$.  
 
\begin{figure}[hbtp]
    \begin{center}
        \includegraphics[width=8.6cm]{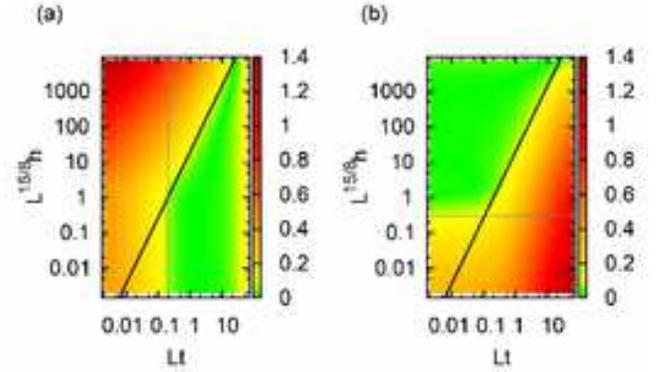}
        \caption{(Color online) Difference between magnetization and its expected scaling behavior about the critical point. 
                 The linear lattice size $L$ was 160.
                (a) $|mL^{1/8}-1.22(Lt)^{1/8}|$ is illustrated. The black line is $t=0.2h^{8/15}$. The vertical gray line  is $Lt=0.2$.
                (b) $|mL^{1/8}-(L^{15/8}h)^{1/15}|$ is illustrated. 
                 The black line is $t=0.2h^{8/15}$. The horizontal gray line is $L^{15/8}h= 0.3$}
        \label{fig:3d_scaledMdiff}
    \end{center}
\end{figure}

Figure \ref{fig:3d_scaledMdiff}(a) and Figure \ref{fig:3d_scaledMdiff} (b) 
show the difference between $\left<|m|\right>L^{1/8}$ and $1.22(Lt)^{1/8}$  and that between $\left<|m|\right>L^{1/8}$ 
and $(L^{15/8}h)^{1/15}$, respectively. 
These data were obtained by the $160\times160$ lattice size simulations.
Note that the factor $1.22$ comes from the exact solution \cite{Yang1952,gaunt1970equation}.  
According to the crossover scaling formalism \cite{fisher1974renormalization}, 
if $t^{-15/8}h$ is large enough, then the magnetization obeys $m\sim t^{1/8}$, 
and if $h^{-8/15}t$ is large enough ($t^{-15/8}h$ is small enough), then it obeys $m\sim h^{1/15}$. 
Figure \ref{fig:3d_scaledMdiff}(a) shows that if the finite-size effects are negligible 
($Lt\gg0.2$) and $t\gg0.2h^{8/15}$ (i.e., $th^{-8/15}$ is large), 
then the critical behavior is $m\sim t^{1/8}$. 
Figure \ref{fig:3d_scaledMdiff}(b) shows that if finite-size effects are negligible 
($L^{15/8}h\gg0.3$) and $t\ll0.2h^{8/15}$ (i.e., $t^{-15/8}h$ is large), 
then the critical behavior is $m\sim h^{1/15}$. 
Thus, Fig.\ \ref{fig:3d_scaledMdiff} clearly shows that the line ($t=0.2h^{8/15}$) gives the boarder of the two scaling regimes. 

\section{Conclusions}

In this work, we introduced a two-dimensional simulated tempering in temperature and external field, which we refer to as 
Simulated Tempering and Magnetizing (STM).
We applied it to the two-dimensional Ising model. 
During the simulations, two-dimensional random walks in temperature 
and external field  were realized. 
The random walk covered a wide area of temperature and external field so that the STM simulations enabled us to study
a wide area of phase diagram from a single simulation run.

Even though the first-order phase transitions along the external field change did not directly occur, the transitions 
happened through high temperature regions, or ``super critical water regions." 
The dimensional extension allowed us to overcome the difficultly of the first-order phase transitions. 
Thus, this result suggests 
that the dimensional extension allows us to overcome the difficulty of crossing the first-order phase transition points with the ST method.
The similarity between ST and REM implies that the dimensional extension of REM will also give this property
(An example is shown for the case of a two-dimensional REM simulation in temperature and pressure in Ref.~\cite{Sugita2001}). 

We also performed STM simulations with several different parameter-updating frequencies. According to the convergence 
and sizes of error bars, the more frequent attempts should be the better choice. The calculated auto-correlation time
also suggested that the frequent attempt is favorable.  

We investigated the crossover behavior of phase transitions by calculating the magnetization per spin $m$ around 
the critical point by the reweighting techniques.      
The results showed agreement with the previous theoretical studies. 
Thus, this supports the validity of the two-dimensional ST method, or STM.

With the data of the present work, we can calculate the two-dimensional density of states $n(E,M)$, 
so that we can determine the weight factor for the two-dimensional multicanonical
simulations.  Therefore, we can also perform the two-dimensional multicanonical simulations. 
The work is in progress.
The STM method will be very useful for simulating spin-glass systems, 
and work is also in progress.
We also remark that the present methods are not only useful for spin systems but also other complex systems with many degrees of freedom.
It should be worth noting that because this method does not modify the energy calculation, 
the method should be very much compatible with existing package programs.

\appendix*

\section{Lattice gas and Ising model}
The total energy of Ising model $H$ on a square lattice can be converted 
into that of lattice gas in the following manner: 
\begin{widetext}
\begin{eqnarray}
H&=&-J\sum_{\left<i,j\right>} \sigma_i \sigma_j -h \sum \sigma_i\\
&=& -J\sum_{\left<i,j\right>} (2s_i-1)(2s_j-1) -h\sum{(2s_i-1)} \,,
\end{eqnarray}
\end{widetext}
where $\sigma_i = \pm 1$ and $s_i=1,0$. If $\sigma_i=1$, then $s_i=1$ and vice versa.
We then have
\begin{widetext}
\begin{eqnarray}
H&=& -4J\sum_{\left<i,j\right>}s_is_j +2J\sum_{\left<i,j\right>}\left(s_i+s_j\right) +J\sum_{\left<i,j\right>}1 -h\sum{(2s_i-1)} \\
&=&-4J\sum_{\left<i,j\right>}s_is_j +8Jn+2JN -2hn+hN \\
&=&-4J\sum_{\left<i,j\right>}s_is_j -(2h-8J)n + (h-2J)N \,,
\end{eqnarray}
\end{widetext}
where $n$ and $N$ are the number of occupied sites and the total number of sites, respectively.
The first term corresponds to the attractive energy between particles of lattice gas. 
The second term corresponds to the chemical potential of lattice gas. 
The last term is a constant.
Here, we define $\mu \equiv (2h-8J)$ and $E_{g}\equiv -4J\sum_{\left<i,j\right>}s_is_j$.

Thus, free energy per spin $f$ is given by
\begin{widetext}
\begin{eqnarray}
\exp (-\beta fN) &=& \sum_{\sigma_0 =\pm1, \sigma_1 =\pm1,\dots,\sigma_N =\pm1 }\exp(-\beta H)\\
				&=&\sum_{s_0 =1,0, s_1 =1,0,\dots,s_N =1,0 } \exp [-\beta(E_{g}-\mu n) ]\exp(-\beta(h-2J)N)\\
				&=&\Theta\exp(-\beta(h-2J)N)\\
				&=&\exp(\beta pN)\exp(-\beta(h-2J)N) \,,
\end{eqnarray}
\end{widetext}
where  $p$ is pressure. 
Instead of $V$, $N$ appears.  The Greek letter $\Theta$ stands for the Grand partition function, where
$\Theta = \sum_{s_0 =1,0, s_1 =1,0,\dots,s_N =1,0 } \exp [-\beta(E_{g}-\mu n) ]$.
 The last two equations were obtained with grand canonical ensembles.
Therefore, we obtain
\begin{eqnarray}
-f &=& p -(h-2J)  \,,\\
p &=& h -f -2J \,.
\end{eqnarray}

Thus, we conclude that the canonical ensemble of Ising model is equivalent to the $\mu$-$T$ ensemble of lattice gas model
with the following correspondence: 
\begin{eqnarray}
p &=& h -f -2J \,,\\
\mu &=& (2h-8J) \,,\\
E_{g}&=& -4J\sum_{\left<i,j\right>}s_is_j .
\end{eqnarray}


\begin{acknowledgments}
We thank Drs.\ Wolfhard Janke, Desmond Johnston, and Thomas Neuhaus for useful discussions.  
Some of the computations were performed on the supercomputers
at the Information Technology Center, Nagoya
University, at the Research Center for Computational
Science, Institute for Molecular Science, and at the Supercomputer Center, 
Institute for Solid State Physics, University of Tokyo. 
This work was supported, in part, by JSPS Institutional Program for Young Researcher Overseas Visit (to T.N.) and 
by Grants-in-Aid for Scientific Research
on Innovative Areas (``Fluctuations and Biological Functions")
and for the Next Generation Super Computing
Project, Nanoscience Program and Computational
Materials Science Initiative from the Ministry of Education,
Culture, Sports, Science and Technology, Japan
(MEXT).

\end{acknowledgments}

\bibliography{citation}

\begin{thebibliography}{37}%
\makeatletter
\providecommand \@ifxundefined [1]{%
 \@ifx{#1\undefined}
}%
\providecommand \@ifnum [1]{%
 \ifnum #1\expandafter \@firstoftwo
 \else \expandafter \@secondoftwo
 \fi
}%
\providecommand \@ifx [1]{%
 \ifx #1\expandafter \@firstoftwo
 \else \expandafter \@secondoftwo
 \fi
}%
\providecommand \natexlab [1]{#1}%
\providecommand \enquote  [1]{``#1''}%
\providecommand \bibnamefont  [1]{#1}%
\providecommand \bibfnamefont [1]{#1}%
\providecommand \citenamefont [1]{#1}%
\providecommand \href@noop [0]{\@secondoftwo}%
\providecommand \href [0]{\begingroup \@sanitize@url \@href}%
\providecommand \@href[1]{\@@startlink{#1}\@@href}%
\providecommand \@@href[1]{\endgroup#1\@@endlink}%
\providecommand \@sanitize@url [0]{\catcode `\\12\catcode `\$12\catcode
  `\&12\catcode `\#12\catcode `\^12\catcode `\_12\catcode `\%12\relax}%
\providecommand \@@startlink[1]{}%
\providecommand \@@endlink[0]{}%
\providecommand \url  [0]{\begingroup\@sanitize@url \@url }%
\providecommand \@url [1]{\endgroup\@href {#1}{\urlprefix }}%
\providecommand \urlprefix  [0]{URL }%
\providecommand \Eprint [0]{\href }%
\providecommand \doibase [0]{http://dx.doi.org/}%
\providecommand \selectlanguage [0]{\@gobble}%
\providecommand \bibinfo  [0]{\@secondoftwo}%
\providecommand \bibfield  [0]{\@secondoftwo}%
\providecommand \translation [1]{[#1]}%
\providecommand \BibitemOpen [0]{}%
\providecommand \bibitemStop [0]{}%
\providecommand \bibitemNoStop [0]{.\EOS\space}%
\providecommand \EOS [0]{\spacefactor3000\relax}%
\providecommand \BibitemShut  [1]{\csname bibitem#1\endcsname}%
\let\auto@bib@innerbib\@empty
\bibitem [{\citenamefont {Hansmann}\ and\ \citenamefont
  {Okamoto}(1999)}]{Hansmann1999}%
  \BibitemOpen
  \bibfield  {author} {\bibinfo {author} {\bibfnamefont {U.~H.~E.}\
  \bibnamefont {Hansmann}}\ and\ \bibinfo {author} {\bibfnamefont
  {Y.}~\bibnamefont {Okamoto}},\ }in\ \href@noop {} {\emph {\bibinfo
  {booktitle} {Annual Reviews of Computational Physics VI}}},\ \bibinfo
  {editor} {edited by\ \bibinfo {editor} {\bibfnamefont {D.}~\bibnamefont
  {Stauffer}}}\ (\bibinfo  {publisher} {World Scientific, Singapore},\ \bibinfo
  {year} {1999})\ pp.\ \bibinfo {pages} {129--157}\BibitemShut {NoStop}%
\bibitem [{\citenamefont {Mitsutake}\ \emph {et~al.}(2001)\citenamefont
  {Mitsutake}, \citenamefont {Sugita},\ and\ \citenamefont
  {Okamoto}}]{Mitsutake2001}%
  \BibitemOpen
  \bibfield  {author} {\bibinfo {author} {\bibfnamefont {A.}~\bibnamefont
  {Mitsutake}}, \bibinfo {author} {\bibfnamefont {Y.}~\bibnamefont {Sugita}}, \
  and\ \bibinfo {author} {\bibfnamefont {Y.}~\bibnamefont {Okamoto}},\ }\href
  {\doibase 10.1002/1097-0282(2001)60:2<96::AID-BIP1007>3.0.CO;2-F} {\bibfield
  {journal} {\bibinfo  {journal} {Biopolymers}\ }\textbf {\bibinfo {volume}
  {60}},\ \bibinfo {pages} {96} (\bibinfo {year} {2001})}\BibitemShut {NoStop}%
\bibitem [{\citenamefont {Sugita}\ and\ \citenamefont
  {Okamoto}(2002)}]{Sugita2001}%
  \BibitemOpen
  \bibfield  {author} {\bibinfo {author} {\bibfnamefont {Y.}~\bibnamefont
  {Sugita}}\ and\ \bibinfo {author} {\bibfnamefont {Y.}~\bibnamefont
  {Okamoto}},\ }in\ \href {\doibase 10.1007/978-3-642-56080-4\_13} {\emph
  {\bibinfo {booktitle} {Lecture Notes in Computational Science and
  Engineering}}},\ \bibinfo {editor} {edited by\ \bibinfo {editor}
  {\bibfnamefont {T.}~\bibnamefont {Schlick}}\ and\ \bibinfo {editor}
  {\bibfnamefont {H.~H.}\ \bibnamefont {Gan}}}\ (\bibinfo  {publisher}
  {Springer},\ \bibinfo {year} {2002})\ pp.\ \bibinfo {pages} {304--332},\
  \bibinfo {note} {http://arxiv.org/abs/cond-mat/0102296}\BibitemShut {NoStop}%
\bibitem [{\citenamefont {Berg}\ and\ \citenamefont
  {Neuhaus}(1991)}]{berg1991multicanonical}%
  \BibitemOpen
  \bibfield  {author} {\bibinfo {author} {\bibfnamefont {B.~A.}\ \bibnamefont
  {Berg}}\ and\ \bibinfo {author} {\bibfnamefont {T.}~\bibnamefont {Neuhaus}},\
  }\href {\doibase 10.1016/0370-2693(91)91256-U} {\bibfield  {journal}
  {\bibinfo  {journal} {Physics Letters B}\ }\textbf {\bibinfo {volume}
  {267}},\ \bibinfo {pages} {249} (\bibinfo {year} {1991})}\BibitemShut
  {NoStop}%
\bibitem [{\citenamefont {Berg}\ and\ \citenamefont
  {Neuhaus}(1992)}]{berg1992multicanonical}%
  \BibitemOpen
  \bibfield  {author} {\bibinfo {author} {\bibfnamefont {B.~A.}\ \bibnamefont
  {Berg}}\ and\ \bibinfo {author} {\bibfnamefont {T.}~\bibnamefont {Neuhaus}},\
  }\href {\doibase 10.1103/PhysRevLett.68.9} {\bibfield  {journal} {\bibinfo
  {journal} {Physical Review Letters}\ }\textbf {\bibinfo {volume} {68}},\
  \bibinfo {pages} {9} (\bibinfo {year} {1992})}\BibitemShut {NoStop}%
\bibitem [{\citenamefont {Lyubartsev}\ \emph {et~al.}(1992)\citenamefont
  {Lyubartsev}, \citenamefont {Martsinovski}, \citenamefont {Shevkunov},\ and\
  \citenamefont {Vorontsov-Velyaminov}}]{Lyubartsev1992}%
  \BibitemOpen
  \bibfield  {author} {\bibinfo {author} {\bibfnamefont {A.~P.}\ \bibnamefont
  {Lyubartsev}}, \bibinfo {author} {\bibfnamefont {A.~A.}\ \bibnamefont
  {Martsinovski}}, \bibinfo {author} {\bibfnamefont {S.~V.}\ \bibnamefont
  {Shevkunov}}, \ and\ \bibinfo {author} {\bibfnamefont {P.~N.}\ \bibnamefont
  {Vorontsov-Velyaminov}},\ }\href {\doibase 10.1063/1.462133} {\bibfield
  {journal} {\bibinfo  {journal} {The Journal of Chemical Physics}\ }\textbf
  {\bibinfo {volume} {96}},\ \bibinfo {pages} {1776} (\bibinfo {year}
  {1992})}\BibitemShut {NoStop}%
\bibitem [{\citenamefont {Marinari}\ and\ \citenamefont
  {Parisi}(1992)}]{marinari1992simulated}%
  \BibitemOpen
  \bibfield  {author} {\bibinfo {author} {\bibfnamefont {E.}~\bibnamefont
  {Marinari}}\ and\ \bibinfo {author} {\bibfnamefont {G.}~\bibnamefont
  {Parisi}},\ }\href {\doibase 10.1209/0295-5075/19/6/002} {\bibfield
  {journal} {\bibinfo  {journal} {Europhysics Letters (EPL)}\ }\textbf
  {\bibinfo {volume} {19}},\ \bibinfo {pages} {451} (\bibinfo {year}
  {1992})}\BibitemShut {NoStop}%
\bibitem [{\citenamefont {Hukushima}\ and\ \citenamefont
  {Nemoto}(1996)}]{hukushima1996exchange}%
  \BibitemOpen
  \bibfield  {author} {\bibinfo {author} {\bibfnamefont {K.}~\bibnamefont
  {Hukushima}}\ and\ \bibinfo {author} {\bibfnamefont {K.}~\bibnamefont
  {Nemoto}},\ }\href {\doibase 10.1143/JPSJ.65.1604} {\bibfield  {journal}
  {\bibinfo  {journal} {Journal of the Physics Society Japan}\ }\textbf
  {\bibinfo {volume} {65}},\ \bibinfo {pages} {1604} (\bibinfo {year}
  {1996})}\BibitemShut {NoStop}%
\bibitem [{\citenamefont {Geyer}(1991)}]{Geyer1991}%
  \BibitemOpen
  \bibfield  {author} {\bibinfo {author} {\bibfnamefont {C.~J.}\ \bibnamefont
  {Geyer}},\ }in\ \href@noop {} {\emph {\bibinfo {booktitle} {Computing Science
  and Statistics, Proceedings of the 23rd Symposium on the Interface}}},\
  \bibinfo {editor} {edited by\ \bibinfo {editor} {\bibfnamefont {E.~M.}\
  \bibnamefont {Keramidas}}}\ (\bibinfo  {publisher} {Interface Foundation of
  North America},\ \bibinfo {year} {1991})\ pp.\ \bibinfo {pages}
  {156--163}\BibitemShut {NoStop}%
\bibitem [{\citenamefont {Wang}\ and\ \citenamefont
  {Landau}(2001{\natexlab{a}})}]{Wang2001a}%
  \BibitemOpen
  \bibfield  {author} {\bibinfo {author} {\bibfnamefont {F.}~\bibnamefont
  {Wang}}\ and\ \bibinfo {author} {\bibfnamefont {D.~P.}\ \bibnamefont
  {Landau}},\ }\href {\doibase 10.1103/PhysRevLett.86.2050} {\bibfield
  {journal} {\bibinfo  {journal} {Physical Review Letters}\ }\textbf {\bibinfo
  {volume} {86}},\ \bibinfo {pages} {2050} (\bibinfo {year}
  {2001}{\natexlab{a}})}\BibitemShut {NoStop}%
\bibitem [{\citenamefont {Wang}\ and\ \citenamefont
  {Landau}(2001{\natexlab{b}})}]{Wang2001b}%
  \BibitemOpen
  \bibfield  {author} {\bibinfo {author} {\bibfnamefont {F.}~\bibnamefont
  {Wang}}\ and\ \bibinfo {author} {\bibfnamefont {D.~P.}\ \bibnamefont
  {Landau}},\ }\href {\doibase 10.1103/PhysRevE.64.056101} {\bibfield
  {journal} {\bibinfo  {journal} {Physical Review E}\ }\textbf {\bibinfo
  {volume} {64}},\ \bibinfo {pages} {056101} (\bibinfo {year}
  {2001}{\natexlab{b}})}\BibitemShut {NoStop}%
\bibitem [{\citenamefont {Laio}\ and\ \citenamefont
  {Parrinello}(2002)}]{Laio2002}%
  \BibitemOpen
  \bibfield  {author} {\bibinfo {author} {\bibfnamefont {A.}~\bibnamefont
  {Laio}}\ and\ \bibinfo {author} {\bibfnamefont {M.}~\bibnamefont
  {Parrinello}},\ }\href {\doibase 10.1073/pnas.202427399} {\bibfield
  {journal} {\bibinfo  {journal} {Proc. Natl. Acad. Sci. USA}\ }\textbf
  {\bibinfo {volume} {99}},\ \bibinfo {pages} {12562} (\bibinfo {year}
  {2002})}\BibitemShut {NoStop}%
\bibitem [{\citenamefont {Swendsen}\ and\ \citenamefont
  {Wang}(1986)}]{Swendsen1986}%
  \BibitemOpen
  \bibfield  {author} {\bibinfo {author} {\bibfnamefont {R.~H.}\ \bibnamefont
  {Swendsen}}\ and\ \bibinfo {author} {\bibfnamefont {J.-S.}\ \bibnamefont
  {Wang}},\ }\href {\doibase 10.1103/PhysRevLett.57.2607} {\bibfield  {journal}
  {\bibinfo  {journal} {Physical Review Letters}\ }\textbf {\bibinfo {volume}
  {57}},\ \bibinfo {pages} {2607} (\bibinfo {year} {1986})}\BibitemShut
  {NoStop}%
\bibitem [{\citenamefont {Iba}(2001)}]{iba2001extended}%
  \BibitemOpen
  \bibfield  {author} {\bibinfo {author} {\bibfnamefont {Y.}~\bibnamefont
  {Iba}},\ }\href {\doibase 10.1142/S0129183101001912} {\bibfield  {journal}
  {\bibinfo  {journal} {International Journal of Modern Physics C}\ }\textbf
  {\bibinfo {volume} {12}},\ \bibinfo {pages} {623} (\bibinfo {year}
  {2001})}\BibitemShut {NoStop}%
\bibitem [{\citenamefont {Kim}\ and\ \citenamefont {Straub}(2010)}]{kim2010}%
  \BibitemOpen
  \bibfield  {author} {\bibinfo {author} {\bibfnamefont {J.}~\bibnamefont
  {Kim}}\ and\ \bibinfo {author} {\bibfnamefont {J.~E.}\ \bibnamefont
  {Straub}},\ }\href {\doibase 10.1063/1.3503503} {\bibfield  {journal}
  {\bibinfo  {journal} {The Journal of Chemical Physics}\ }\textbf {\bibinfo
  {volume} {133}},\ \bibinfo {pages} {154101} (\bibinfo {year}
  {2010})}\BibitemShut {NoStop}%
\bibitem [{\citenamefont {Mitsutake}\ and\ \citenamefont
  {Okamoto}(2009{\natexlab{a}})}]{Mitsutake2009multidimensional1}%
  \BibitemOpen
  \bibfield  {author} {\bibinfo {author} {\bibfnamefont {A.}~\bibnamefont
  {Mitsutake}}\ and\ \bibinfo {author} {\bibfnamefont {Y.}~\bibnamefont
  {Okamoto}},\ }\href {\doibase 10.1103/PhysRevE.79.047701} {\bibfield
  {journal} {\bibinfo  {journal} {Physical Review E}\ }\textbf {\bibinfo
  {volume} {79}},\ \bibinfo {pages} {047701} (\bibinfo {year}
  {2009}{\natexlab{a}})}\BibitemShut {NoStop}%
\bibitem [{\citenamefont {Mitsutake}\ and\ \citenamefont
  {Okamoto}(2009{\natexlab{b}})}]{Mitsutake2009multidimensional2}%
  \BibitemOpen
  \bibfield  {author} {\bibinfo {author} {\bibfnamefont {A.}~\bibnamefont
  {Mitsutake}}\ and\ \bibinfo {author} {\bibfnamefont {Y.}~\bibnamefont
  {Okamoto}},\ }\href {\doibase 10.1063/1.3127783} {\bibfield  {journal}
  {\bibinfo  {journal} {The Journal of Chemical Physics}\ }\textbf {\bibinfo
  {volume} {130}},\ \bibinfo {pages} {214105} (\bibinfo {year}
  {2009}{\natexlab{b}})}\BibitemShut {NoStop}%
\bibitem [{\citenamefont {Mitsutake}(2009)}]{Mitsutake2009MSTMREM}%
  \BibitemOpen
  \bibfield  {author} {\bibinfo {author} {\bibfnamefont {A.}~\bibnamefont
  {Mitsutake}},\ }\href {\doibase 10.1063/1.3204443} {\bibfield  {journal}
  {\bibinfo  {journal} {The Journal of Chemical Physics}\ }\textbf {\bibinfo
  {volume} {131}},\ \bibinfo {pages} {094105} (\bibinfo {year}
  {2009})}\BibitemShut {NoStop}%
\bibitem [{\citenamefont {Gaunt}\ and\ \citenamefont
  {Domb}(1970)}]{gaunt1970equation}%
  \BibitemOpen
  \bibfield  {author} {\bibinfo {author} {\bibfnamefont {D.~S.}\ \bibnamefont
  {Gaunt}}\ and\ \bibinfo {author} {\bibfnamefont {C.}~\bibnamefont {Domb}},\
  }\href {\doibase 10.1088/0022-3719/3/7/003} {\bibfield  {journal} {\bibinfo
  {journal} {Journal of Physics C: Solid State Physics}\ }\textbf {\bibinfo
  {volume} {3}},\ \bibinfo {pages} {1442} (\bibinfo {year} {1970})}\BibitemShut
  {NoStop}%
\bibitem [{\citenamefont {Chodera}\ and\ \citenamefont
  {Shirts}(2011)}]{chodera2011replica}%
  \BibitemOpen
  \bibfield  {author} {\bibinfo {author} {\bibfnamefont {J.~D.}\ \bibnamefont
  {Chodera}}\ and\ \bibinfo {author} {\bibfnamefont {M.~R.}\ \bibnamefont
  {Shirts}},\ }\href {\doibase 10.1063/1.3660669} {\bibfield  {journal}
  {\bibinfo  {journal} {The Journal of Chemical Physics}\ }\textbf {\bibinfo
  {volume} {135}},\ \bibinfo {pages} {194110} (\bibinfo {year}
  {2011})}\BibitemShut {NoStop}%
\bibitem [{\citenamefont {Zhang}\ and\ \citenamefont {Ma}(2010)}]{zhang2010}%
  \BibitemOpen
  \bibfield  {author} {\bibinfo {author} {\bibfnamefont {C.}~\bibnamefont
  {Zhang}}\ and\ \bibinfo {author} {\bibfnamefont {J.}~\bibnamefont {Ma}},\
  }\href {\doibase 10.1063/1.3435332} {\bibfield  {journal} {\bibinfo
  {journal} {The Journal of Chemical Physics}\ }\textbf {\bibinfo {volume}
  {132}},\ \bibinfo {pages} {244101} (\bibinfo {year} {2010})}\BibitemShut
  {NoStop}%
\bibitem [{\citenamefont {Matsumoto}\ and\ \citenamefont
  {Nishimura}(1998)}]{matsumoto1998mersenne}%
  \BibitemOpen
  \bibfield  {author} {\bibinfo {author} {\bibfnamefont {M.}~\bibnamefont
  {Matsumoto}}\ and\ \bibinfo {author} {\bibfnamefont {T.}~\bibnamefont
  {Nishimura}},\ }\href@noop {} {\bibfield  {journal} {\bibinfo  {journal} {ACM
  Transactions on Modeling and Computer Simulation (TOMACS)}\ }\textbf
  {\bibinfo {volume} {8}},\ \bibinfo {pages} {3} (\bibinfo {year}
  {1998})}\BibitemShut {NoStop}%
\bibitem [{\citenamefont {Mitsutake}\ and\ \citenamefont
  {Okamoto}(2000)}]{mitsutake2000replica}%
  \BibitemOpen
  \bibfield  {author} {\bibinfo {author} {\bibfnamefont {A.}~\bibnamefont
  {Mitsutake}}\ and\ \bibinfo {author} {\bibfnamefont {Y.}~\bibnamefont
  {Okamoto}},\ }\href {\doibase
  http://dx.doi.org/10.1016/S0009-2614(00)01262-8} {\bibfield  {journal}
  {\bibinfo  {journal} {Chemical Physics Letters}\ }\textbf {\bibinfo {volume}
  {332}},\ \bibinfo {pages} {131} (\bibinfo {year} {2000})}\BibitemShut
  {NoStop}%
\bibitem [{\citenamefont {Ferrenberg}\ and\ \citenamefont
  {Swendsen}(1989)}]{Ferrenberg1989}%
  \BibitemOpen
  \bibfield  {author} {\bibinfo {author} {\bibfnamefont {A.}~\bibnamefont
  {Ferrenberg}}\ and\ \bibinfo {author} {\bibfnamefont {R.}~\bibnamefont
  {Swendsen}},\ }\href {\doibase 10.1103/PhysRevLett.63.1195} {\bibfield
  {journal} {\bibinfo  {journal} {Physical Review Letters}\ }\textbf {\bibinfo
  {volume} {63}},\ \bibinfo {pages} {1195} (\bibinfo {year}
  {1989})}\BibitemShut {NoStop}%
\bibitem [{\citenamefont {Kumar}\ \emph {et~al.}(1992)\citenamefont {Kumar},
  \citenamefont {Rosenberg}, \citenamefont {Bouzida}, \citenamefont
  {Swendsen},\ and\ \citenamefont {Kollman}}]{kumar1992weighted}%
  \BibitemOpen
  \bibfield  {author} {\bibinfo {author} {\bibfnamefont {S.}~\bibnamefont
  {Kumar}}, \bibinfo {author} {\bibfnamefont {J.~M.}\ \bibnamefont
  {Rosenberg}}, \bibinfo {author} {\bibfnamefont {D.}~\bibnamefont {Bouzida}},
  \bibinfo {author} {\bibfnamefont {R.~H.}\ \bibnamefont {Swendsen}}, \ and\
  \bibinfo {author} {\bibfnamefont {P.~A.}\ \bibnamefont {Kollman}},\ }\href
  {\doibase 10.1002/jcc.540130812} {\bibfield  {journal} {\bibinfo  {journal}
  {Journal of Computational Chemistry}\ }\textbf {\bibinfo {volume} {13}},\
  \bibinfo {pages} {1011} (\bibinfo {year} {1992})}\BibitemShut {NoStop}%
\bibitem [{\citenamefont {Shirts}\ and\ \citenamefont
  {Chodera}(2008)}]{shirts2008statistically}%
  \BibitemOpen
  \bibfield  {author} {\bibinfo {author} {\bibfnamefont {M.~R.}\ \bibnamefont
  {Shirts}}\ and\ \bibinfo {author} {\bibfnamefont {J.~D.}\ \bibnamefont
  {Chodera}},\ }\href {\doibase 10.1063/1.2978177} {\bibfield  {journal}
  {\bibinfo  {journal} {The Journal of Chemical Physics}\ }\textbf {\bibinfo
  {volume} {129}},\ \bibinfo {pages} {124105} (\bibinfo {year}
  {2008})}\BibitemShut {NoStop}%
\bibitem [{\citenamefont {Mitsutake}\ \emph {et~al.}(2003)\citenamefont
  {Mitsutake}, \citenamefont {Sugita},\ and\ \citenamefont
  {Okamoto}}]{Mitsutake2003}%
  \BibitemOpen
  \bibfield  {author} {\bibinfo {author} {\bibfnamefont {A.}~\bibnamefont
  {Mitsutake}}, \bibinfo {author} {\bibfnamefont {Y.}~\bibnamefont {Sugita}}, \
  and\ \bibinfo {author} {\bibfnamefont {Y.}~\bibnamefont {Okamoto}},\ }\href
  {\doibase 10.1063/1.1555847} {\bibfield  {journal} {\bibinfo  {journal} {The
  Journal of Chemical Physics}\ }\textbf {\bibinfo {volume} {118}},\ \bibinfo
  {pages} {6664} (\bibinfo {year} {2003})}\BibitemShut {NoStop}%
\bibitem [{\citenamefont {Binder}(1981)}]{binder1981finite}%
  \BibitemOpen
  \bibfield  {author} {\bibinfo {author} {\bibfnamefont {K.}~\bibnamefont
  {Binder}},\ }\href {\doibase 10.1007/BF01293604} {\bibfield  {journal}
  {\bibinfo  {journal} {Zeitschrift f{\"u}r Physik B Condensed Matter}\
  }\textbf {\bibinfo {volume} {43}},\ \bibinfo {pages} {119} (\bibinfo {year}
  {1981})}\BibitemShut {NoStop}%
\bibitem [{\citenamefont {Miller}(1974)}]{Miller1974}%
  \BibitemOpen
  \bibfield  {author} {\bibinfo {author} {\bibfnamefont {R.~G.}\ \bibnamefont
  {Miller}},\ }\href {\doibase 10.1093/biomet/61.1.1} {\bibfield  {journal}
  {\bibinfo  {journal} {Biometrika}\ }\textbf {\bibinfo {volume} {61}},\
  \bibinfo {pages} {1} (\bibinfo {year} {1974})}\BibitemShut {NoStop}%
\bibitem [{\citenamefont {Berg}(2004)}]{berg2004book}%
  \BibitemOpen
  \bibfield  {author} {\bibinfo {author} {\bibfnamefont {B.~A.}\ \bibnamefont
  {Berg}},\ }\href@noop {} {\emph {\bibinfo {title} {Markov Chain Monte Carlo
  Simulations and Their Statistical Analysis}}}\ (\bibinfo  {publisher} {World
  Scientific, Singapore},\ \bibinfo {year} {2004})\ \bibinfo {note} {;
  http://www.worldscibooks.com/physics/5602.html}\BibitemShut {NoStop}%
\bibitem [{\citenamefont {Lee}\ and\ \citenamefont
  {Yang}(1952)}]{lee1952statistical}%
  \BibitemOpen
  \bibfield  {author} {\bibinfo {author} {\bibfnamefont {T.~D.}\ \bibnamefont
  {Lee}}\ and\ \bibinfo {author} {\bibfnamefont {C.~N.}\ \bibnamefont {Yang}},\
  }\href {\doibase 10.1103/PhysRev.87.410} {\bibfield  {journal} {\bibinfo
  {journal} {Physical Review}\ }\textbf {\bibinfo {volume} {87}},\ \bibinfo
  {pages} {410} (\bibinfo {year} {1952})}\BibitemShut {NoStop}%
\bibitem [{\citenamefont {Ferdinand}\ and\ \citenamefont
  {Fisher}(1969)}]{Ferdinand1969IsingFinite}%
  \BibitemOpen
  \bibfield  {author} {\bibinfo {author} {\bibfnamefont {A.~E.}\ \bibnamefont
  {Ferdinand}}\ and\ \bibinfo {author} {\bibfnamefont {M.~E.}\ \bibnamefont
  {Fisher}},\ }\href {\doibase 10.1103/PhysRev.185.832} {\bibfield  {journal}
  {\bibinfo  {journal} {Phys. Rev.}\ }\textbf {\bibinfo {volume} {185}},\
  \bibinfo {pages} {832} (\bibinfo {year} {1969})}\BibitemShut {NoStop}%
\bibitem [{\citenamefont {Sindhikara}\ \emph {et~al.}(2008)\citenamefont
  {Sindhikara}, \citenamefont {Meng},\ and\ \citenamefont
  {Roitberg}}]{sindhikara2008exchange}%
  \BibitemOpen
  \bibfield  {author} {\bibinfo {author} {\bibfnamefont {D.~J.}\ \bibnamefont
  {Sindhikara}}, \bibinfo {author} {\bibfnamefont {Y.}~\bibnamefont {Meng}}, \
  and\ \bibinfo {author} {\bibfnamefont {A.~E.}\ \bibnamefont {Roitberg}},\
  }\href {\doibase 10.1063/1.2816560} {\bibfield  {journal} {\bibinfo
  {journal} {The Journal of Chemical Physics}\ }\textbf {\bibinfo {volume}
  {128}},\ \bibinfo {pages} {024103} (\bibinfo {year} {2008})}\BibitemShut
  {NoStop}%
\bibitem [{\citenamefont {Sindhikara}\ \emph {et~al.}(2010)\citenamefont
  {Sindhikara}, \citenamefont {Emerson},\ and\ \citenamefont
  {Roitberg}}]{sindhikara2010}%
  \BibitemOpen
  \bibfield  {author} {\bibinfo {author} {\bibfnamefont {D.~J.}\ \bibnamefont
  {Sindhikara}}, \bibinfo {author} {\bibfnamefont {D.~J.}\ \bibnamefont
  {Emerson}}, \ and\ \bibinfo {author} {\bibfnamefont {A.~E.}\ \bibnamefont
  {Roitberg}},\ }\href {\doibase 10.1021/ct100281c} {\bibfield  {journal}
  {\bibinfo  {journal} {Journal of Chemical Theory and Computation}\ }\textbf
  {\bibinfo {volume} {6}},\ \bibinfo {pages} {2804} (\bibinfo {year}
  {2010})}\BibitemShut {NoStop}%
\bibitem [{\citenamefont {Landau}(1976)}]{PhysRevB.13.2997}%
  \BibitemOpen
  \bibfield  {author} {\bibinfo {author} {\bibfnamefont {D.~P.}\ \bibnamefont
  {Landau}},\ }\href {\doibase 10.1103/PhysRevB.13.2997} {\bibfield  {journal}
  {\bibinfo  {journal} {Physical Review B}\ }\textbf {\bibinfo {volume} {13}},\
  \bibinfo {pages} {2997} (\bibinfo {year} {1976})}\BibitemShut {NoStop}%
\bibitem [{\citenamefont {Yang}(1952)}]{Yang1952}%
  \BibitemOpen
  \bibfield  {author} {\bibinfo {author} {\bibfnamefont {C.~N.}\ \bibnamefont
  {Yang}},\ }\href {\doibase 10.1103/PhysRev.85.808} {\bibfield  {journal}
  {\bibinfo  {journal} {Physical Review}\ }\textbf {\bibinfo {volume} {85}},\
  \bibinfo {pages} {808} (\bibinfo {year} {1952})}\BibitemShut {NoStop}%
\bibitem [{\citenamefont {Fisher}(1974)}]{fisher1974renormalization}%
  \BibitemOpen
  \bibfield  {author} {\bibinfo {author} {\bibfnamefont {M.~E.}\ \bibnamefont
  {Fisher}},\ }\href {\doibase 10.1103/RevModPhys.46.597} {\bibfield  {journal}
  {\bibinfo  {journal} {Reviews of Modern Physics}\ }\textbf {\bibinfo {volume}
  {46}},\ \bibinfo {pages} {597} (\bibinfo {year} {1974})}\BibitemShut
  {NoStop}%
\end{thebibliography}%

\end{document}